\documentclass[preprint,showpacs,preprintnumbers,amsmath,amssymb]{revtex4}
\usepackage{graphicx}
\usepackage{dcolumn}
\usepackage{bm}
\usepackage{rotating}
\usepackage{natbib}
\usepackage{txfonts}

\begin{document}

  \title{ Temperature and density dependence of asymmetric nuclear matter and protoneutron star properties within an extended relativistic mean field model}
  \author{Gulshan Mahajan$^{a,b}$}
  \email{gul.mahajan@yahoo.co.in}
  \author{ Shashi K. Dhiman$^{a,c}$}
  \email{shashi.dhiman@gmail.com}

   \affiliation{
$^a$ Department of Physics, Himachal Pradesh University, Shimla - 171005, India. \\
$^b$ Department of Physics, R.G.M. Government College Joginder Nagar - 175015, India.\\
$^c$ University Institute of Natural Sciences and Interface Technologies, Himachal Pradesh Technical University, Post Box  12, Hamirpur, Pin 177001, India.}
\begin{abstract}
The effect of temperature and density dependence of  the
asymmetric nuclear matter properties is studied within the
extended relativistic mean field (ERMF) model, which includes the
contribution from the self and mixed interaction terms by using
different parametrizations obtained by varying the neutron skin
thickness $\Delta$r and $\omega$-meson self-coupling ($\zeta$). We
observed that the symmetry energy and its slope and
incompressibility coefficients  decrease with increasing
temperatures up to saturation densities. The ERMF parametrizations
were employed to obtain a new set of equations of state (EOS) of
the protoneutron star (PNS) with and without inclusion of
hyperons. In our calculations, in comparison with cold compact
stars, we obtained that the gravitational mass of the protoneutron
star with and without hyperons increased by  $\sim 0.4M_{\odot}$
and its radius increased by $\sim 3$km. Whereas in case of the
rotating PNS, the mass shedding limit decreased with increasing
temperature, and this suggested that the keplerian frequency of
the PNS, at T = 10 MeV should be smaller by $14-20\%$ for the EOS
with hyperon, as compared to the keplerian frequency of a cold
compact star.
\end{abstract}


\maketitle

\section{Introduction}

The behavior of nuclear matter at high density and finite
temperature is one  of the challenging problems in contemporary
modern nuclear physics. Among the successful and widely used
approaches to study nuclear matter are nonrelativistic mean field
theory, with effective nucleon-nucleon interactions such as skyrme
forces \cite{Chabanat97,Stone03,Agrawal06}, and relativistic mean
field (RMF) theory \cite{Steiner05}. The RMF theory is more
fundamental as it starts from hadronic field theory with strongly
interacting baryons and mesons as degrees of freedom
\cite{Walecka74}, and it describes very well the basic properties
of nuclei near the valley of stability \cite{Ring96} and the
properties of exotic nuclei with large numbers of neutrons or
protons \cite{Meng96}. The properties of cold nuclear matter can
be studied by imposing the constraints of bulk nuclear matter
properties at the saturation density $\rho_0 = 0.16 fm^{-3}$,
recent experimental limits establish the following values:
symmetry energy $E_{sym}=30 \pm 5 MeV $ \cite{Li2005,Li2007},
slope of symmetry energy $L=88 \pm 25 MeV$ \cite{Li08}, and
incompressibility coefficient $K=240 \pm 20 MeV$
\cite{Li07,Youngblood99,Ma02}. It is considered theoretically that
the density dependence of symmetry energy can be represented by
$E_{sym}(\rho_0)=31.6(\rho/\rho_0)^\gamma$, with
$\gamma$=0.69-1.05 at subnormal density \cite{Li08}, which led to
the extraction of a value for the slope of the nuclear symmetry of
energy $L=88 \pm 25 MeV$. This symmetry energy value is also in
harmony with the symmetry energy obtained from the isoscaling
analysis of the isotope ratio in intermediate energy heavy ion
collisions \cite{Shetty07}.

Recently, heavy-ion reactions induced in laboratories have
provided the  atmosphere necessary to produce hot neutron rich
matter similar to that existing in astrophysical situations. The
reactions especially, which are induced by radioactive beams,
provide a unique means to investigate the isospin-dependent
properties of asymmetric nuclear matter at Cooling Storage Ring
(CSR) at the HIRFL in China, the Radioactive Ion Beam (RIB) at
RIKEN in Japan, FAIR/GSI in Germany, SPIRAL2/GANIL in France, the
Facility for Rare Isotope Beam (FRIB) in the United States. The
heavy ion collision data from analyzing isospin diffusion and size
of the neutron skin in $^{208}Pb$ \cite{Steiner05,Li08} have
helped us significanty in understanding symmetry energy. The
symmetry energy of hot neutron rich matter in a low density regime
\cite{Natowitz10} is important for understanding the liquid gas
phase transition of asymmetric nuclear matter, the dynamical
evolution of massive stars and the supernova explosion mechanism.

The density dependence of symmetry energy influences the nature
and stability of the phases of compact star (CS), the feasibility
of direct URCA cooling process within interior of CS, the
composition and thickness of inner crust of CS, the frequency of
its crustal vibrations and radius of CS. Many correlations have
been studied to understand the density behavior of symmetry energy
\cite{Tsang09}. However, the fundamental origin of this apparent
evolution of symmetry energy is still not clear, and it is
particularly important to understand to what degree its evolution
depends on the density and /or temperature of nuclear matter.
Apart from symmetry energy, the nuclear matter equation of state
(EOS) also depend upon the values of incompressibility
\cite{Yoshida98}. In recent times the giant monopole resonance has
made it possible to find the value of incompressibility
\cite{Li10}. Accurate knowledge of the density and temperature
dependence of  symmetry energy and incompressibility can lead to
plausible EOS of the asymmetric matter.

The properties of the compact stars are mainly determined by the
EOS of nuclear dense matter, which is charge neutral matter in
$\beta$ equilibrium \cite{Shen02}. Any given EOS of baryonic
matter determines  uniquely the mass-radius relationship of a
compact star and, in particular, the maximum mass a  compact star
can achieve before collapsing into a black hole \cite{Schulze06}.
Theoretical investigations of high-density $\beta$ stable have
lead to the conclusion that hyperons will appear at densities of
about 2-4 times the saturation density $(\rho_0)$ and soften the
EOS in high density regimes, as the conversion of nucleons to
hyperons can relieve the fermi surface and leads to a reduction of
compact star mass \cite{Shapiro83,Glendenning00}. For compact star
matter with uniform distribution, the
composition is determined by the requirement of charge neutrality and $\beta$%
-equilibrium conditions. The threshold density for a hyperon
species is determined not only by its charge and mass but also by
the meson fields. The stiffer the EOS without hyperons is, the
greater is the softening effect when hyperons are included
\cite{Balberg99,Schulze06}. Further the presence of hyperons
should allow direct URCA like cooling involving the beta decay of
the hyperons.

In this work we have extended our previous ERMF model
\cite{Dhiman07} to study the effect of temperature on asymmetric
nuclear matter and the properties of  protoneutron stars (PNS).
This paper is organized as follows. In Sec. II, we briefly
describe the extended relativistic mean field theory. In Sec. III
we present the results and discussion for nuclear matter
properties and structure properties of non rotating PNS and
rotating PNS with keplerian frequency. In Sec. IV we present our
conclusions for the present work.

\section{Formalism}
The Lagrangian density for the ERMF model describes the interactions
 from self and mixed terms for the scalar-isoscalar ($\sigma$),
vector-isoscalar ($\omega$), and vector-isovector ($\rho$) mesons
\cite{Dhiman07,Serot97}. For completeness the Lagrangian density
for the extended ERMF model can be written as,
\begin{equation}
\label{eq:lden}
{\cal L}= {\cal L_{BM}}+{\cal L_{\sigma}} + {\cal L_{\omega}} + {\cal L_{\mathbf{\rho}}} + {\cal L_{\sigma\omega\mathbf{\rho}}}.
\end{equation}
The description of the various terms of the Lagrangian  and the
Euler-Lagrangian equations for ground state expectation values of
the meson fields are provided in Ref. \cite{Dhiman07}. At finite
temperatures the  baryon vector density $\rho_B$,  scalar density
$\rho_{sB}$ and charge density $\rho_{p}$ are, respectively,
\begin{equation}
\rho_{B}= \left< \overline{\Psi}_B \gamma^0 \Psi_B\right>(n_i-\overline{n_i}) = \frac{\gamma}{(2\pi)^{3}}\int_{0}^{k_{B}}d^{3}k (n_i-\overline{n_i}),
\end{equation}

\begin{equation}
\rho_{sB} = \left< \overline{\Psi}_B\Psi_B \right>(n_i+\overline{n_i})
          = \frac{\gamma}{(2\pi)^3}\int_{0}^{k_{B}}d^{3}k \frac{M_{B}^*}
            {\sqrt{k^2 + M_{B}^{*2}}}(n_i+\overline{n_i}),
\end{equation}
\begin{equation}
\rho_{p} = \left< \overline{\Psi}_B\gamma^{0}\frac{1+\tau_{3B}}{2}\Psi_B \right>(n_i+\overline{n_i}).
\end{equation}
 Here, $\gamma$ is the spin-isospin degeneracy.  $M_{B}^{*} = M_{B} - g_{\sigma B}\sigma - g_{\sigma^{*}B}\sigma ^{*}$
is the effective mass of the baryon species B = (p, n, $\Lambda,
\Sigma^{\pm}, \Xi^{\pm}$), $k_{B}$ is its Fermi momentum and
$\tau_{3B}$ denotes the isospin projections of baryon B.

The thermal distribution function in these expression are defined by
\begin{equation}
n_i=\frac{1}{e^{\beta(\epsilon^*_i-\nu_i)}+1} \qquad
\overline{n_i}=\frac{1}{e^{\beta(\epsilon^*_i+\nu_i)}+1}
\end{equation}

where
\begin{equation}
\epsilon^*_i=\sqrt{k^2+M^{*2}_B}\qquad
\nu_i=\mu-g_{\omega N}\omega\pm g_{\rho} \frac{\rho}{2} \qquad
 (i = n, p)
\end{equation}
are the effective energy and effective chemical potential
respectively.

The energy density of the uniform matter  in the ERMF models is given by
\begin{equation}
\label{eq:eden}
\begin{split}
{\cal E} & = \sum_{j=p,n}\frac{1}{\pi^{2}}\int_{0}^{k_j}
\epsilon_{i}^{*}(k) k^2 dk (n_i+\overline{n_i}) +\sum_{B}g_{\omega
B}\omega\rho_{B}+\sum_{B}g_{\rho B}\tau_{3B}\rho
+ \frac{1}{2}m_{\sigma}^2\sigma^2\\
&+\frac{\overline{\kappa}}{6}g_{\sigma N}^3\sigma^3
+\frac{\overline{\lambda}}{24}g_{\sigma N}^4\sigma^4
-\frac{\zeta}{24}g_{\omega N}^4\omega^4-\frac{\xi}{24}g_{\rho N}^4\rho^4
 - \frac{1}{2} m_{\omega}^2 \omega ^2
-\frac{1}{2} m_{\rho}^2 \rho ^2\\
&-\overline{\alpha_1} g_{\sigma N}
 g_{\omega N}^{2}\sigma \omega^2-\frac{1}{2}
\overline{\alpha_1}^\prime g_{\sigma N}^2 g_{\omega N}^2\sigma^2 \omega^2
-\overline{\alpha_2}g_{\sigma N}g_{\rho N}^2 \sigma\rho^2
 -\frac{1}{2} \overline{\alpha_2}^\prime g_{\sigma N}^2 g_{\rho N}^2\sigma^2
\rho^2\\
& - \frac{1}{2} \overline{\alpha_3}^\prime g_{\omega N}^2 g_{\rho N}^2
\omega^2\rho^2 +\frac{1}{2} m_{\sigma^*}^{2} {\sigma ^*} ^{2}+\sum_{B}g_{\phi B}\phi\rho_{B} -\frac{1}{2} m_{\phi}^{2} {\phi} ^{2}.
\end{split}
\end{equation}
The pressure of the uniform matter  is given by
\begin{equation}
\label{eq:pden}
\begin{split}
P & = \sum_{j=p,n}\frac{1}{3\pi^{2}}\int_{0}^{k_j}
\frac{k^{4}dk}{\sqrt{k^2+M_{j}^{*2}}}(n_i+\overline{n_i})
- \frac{1}{2}m_{\sigma}^2\sigma^2-\frac{\overline{\kappa}}{6}g_{\sigma N}^3\sigma^3 -\frac{\overline{\lambda}}{24}g_{\sigma N}^4\sigma^4\\
& +\frac{\zeta}{24}g_{\omega N}^4\omega^4+\frac{\xi}{24}g_{\rho N}^4\rho^4
  + \frac{1}{2} m_{\omega}^2 \omega ^2
+\frac{1}{2} m_{\rho}^2 \rho ^2+\overline{\alpha_1} g_{\sigma N}
g_{\omega N}^{2}\sigma \omega^2\\
&+\frac{1}{2} \overline{\alpha_1}^\prime g_{\sigma N}^2 g_{\omega N}^2\sigma^2 \omega^2+\overline{\alpha_2}g_{\sigma N}g_{\rho N}^2 \sigma\rho^2
 +\frac{1}{2} \overline{\alpha_2}^\prime g_{\sigma N}^2 g_{\rho N}^2\sigma^2
\rho^2\\
& + \frac{1}{2} \overline{\alpha_3}^\prime g_{\omega N}^2 g_{\rho N}^2
\omega^2\rho^2
 -\frac{1}{2} m_{\sigma^*}^{2} {\sigma ^*} ^{2}
 +\frac{1}{2} m_{\phi}^{2} {\phi} ^{2}.%
\end{split}
\end{equation}
The symmetry energy $E_{sym}$, the slope $L$, and the
incompressibility $K$ can be evaluated as
\begin{equation}
E_{sym}(\rho)=\frac{1}{2} \frac{d^2E(\rho,\delta)}{d\delta^2}\Bigg|_{\delta=0}
\end{equation}
\begin{equation}
L=3\rho_0\frac{dE_{sym}(\rho)}{d\rho}\Bigg|_{\delta=\rho_0}
\end{equation}
\begin{equation}
K=9\rho_0^2\frac{d^2E_0(\rho)}{d\rho^2}\Bigg|_{\rho=\rho_0}
\end{equation}
where $\rho_0$ is the saturation density, $E(\rho.\delta)$ is the energy per
nucleons at a given density $\rho$ and asymmetry parameter
 $\delta = (\frac{\rho_n-\rho_p}{\rho_n+\rho_p})$ and  $E_0(\rho)=E(\rho,\delta=0)$ is the energy per nucleon for symmetric matter.

\section{Result and discussions}

In the present work we have employed parametrization sets of the
ERMF model,
 BSR1 - BSR21 \cite{Dhiman07,Agrawal10}, generated by varying
the $\omega$ meson self-coupling $\zeta$ and neutron skin
thickness $\Delta$r for the $^{208}Pb$ nucleus. These
parametrizations have been obtained so as to reproduce the nuclear
structure properties  in finite nuclei and bulk properties of
nuclear matter at nuclear saturation density \cite{Dhiman07}. The
parametrization sets BSR1-BSR7 correspond to the value of $\omega$
meson self-coupling $\zeta$ = 0.0, sets BSR8-BSR14 correspond to
$\zeta$ = 0.03, and sets BSR15-BSR21 correspond to $\zeta$ = 0.06,
and for each parametrization set the value of neutron skin
thickness of $^{208}$Pb varies from 0.16 to 0.28 fm in  intervals
of 0.02 fm. Further, the hyperon-meson coupling parameters are
expressed in terms of the nucleon-meson coupling using the SU(6)
model. The coupling parameters of $\sigma$-meson-hyperon and
$\omega$-meson-hyperon are very sensitive to structural properties
of compact stars, so these parameters have been fitted to the
hyperon-nucleon potential depth the same as in Ref.
\cite{Dhiman07}, and its value $X_{\omega y}$ varies from 0.5 to
0.8, where $X_{\omega y}$ is defined as,
\begin{eqnarray}
X_{\omega Y}=\left\{
\begin{array}{cc}
\left( \frac{g_{\omega Y}}{g_{\omega N}}\right)&   \hspace{0.4cm}for \hspace{0.4cm}\Lambda \hspace{0.4cm} and \hspace{0.2cm} \Sigma \hspace{0.2cm}  hyperons\\
2\left( \frac{g_{\omega Y}}{g_{\omega N}}\right)&   for\hspace{0.4cm} \Xi\hspace{0.4cm}  hyperons,
\end{array}
\right .
\label{eq:xw}
\end{eqnarray}
where $g_{\omega Y}$ and $g_{\omega N}$ are the $\omega$-
meson-hyperon and $\omega$-meson-nucleon coupling parameters.
\subsection{Nuclear Matter Properties}

We study the properties of symmetric and asymmetric nuclear matter
for different parametrizations of the ERMF model at temperatures
of 0 to 30 MeV. In Table \ref{tab1} we present the results for the
bulk properties of nuclear matter at saturation density for the
parameters BSR1, BSR7, BSR8, BSR14, BSR15, and BSR21 at
temperatures T = 0, 10, 20, and 30 MeV. It is found that the bulk
properties at saturation densities remain almost the same up to 20
MeV, but as temperature increases further these properties start
varying significantly. The results for the saturation density
($\rho_0$), energy per nucleon (E/A), incompressibility
coefficient for symmetric nuclear matter (K), symmetry energy
($E_{sym}(\rho_0)$), linear density dependence of symmetry energy
slope (L) and effective nucleon mass ($M^*$) for the various
parametrizations at saturation density are given in Table
\ref{tab1}. It can be seen from  Table \ref{tab1} that the nuclear
matter properties at saturation density such as energy per
nucleon, symmetry energy and value of its slope, and the effective
mass of the nucleons get changed beyond T $\geq$ 20 MeV by a very
small amount with respect to T = 0 MeV for the all
parametrizations of the ERMF model. However,  the
incompressibility coefficient for symmetric nuclear matter
decreases up to a maximum of $12.5 \%$ at T= 30 MeV with respect
to T = 0 MeV for the BSR1 parametrization, which provides the
stiffest EOS with neutron star gravitational mass M = $2.5
M_{\odot}$ \cite{Dhiman07}. The variation in the values of K is a
minimum of $7\%$ for the BSR21 parametrization, which provides the
softest EOS with neutron star gravitational mass M =
$1.74M_{\odot}$ \cite{Dhiman07}.

\begin{figure}
 \includegraphics[width=15cm,angle=270]{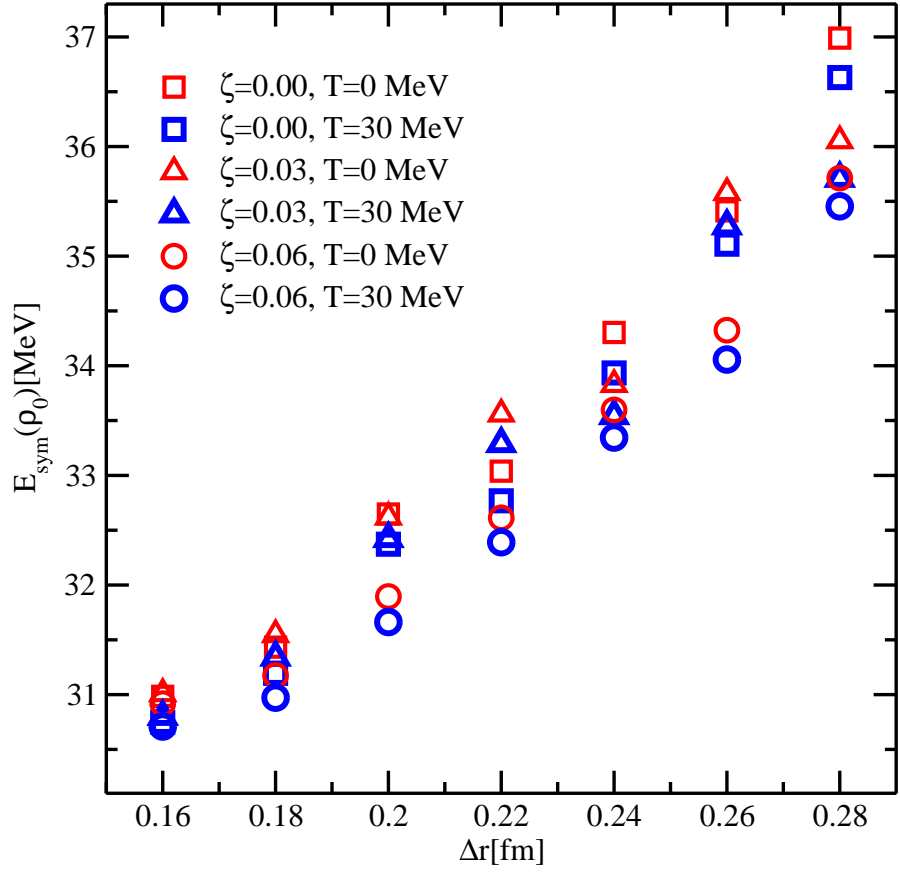}
  \caption{\label{fig:fig1}
(Color online) The symmetry energy $E_{sym}(\rho_0)$ plotted as a
function of the neutron skin thickness $\Delta$r in the $^{208}Pb$
nucleus for 21 different parametrizations of the ERMF model. The
squares, triangles and circles represent results for the
parametrizations BSR1-BSR7, BSR8-BSR14 and BSR15-BSR21
respectively. The red symbols represent the results at T = 0 MeV
and the blue symbols represent the results at T = 30 MeV. }
   \end{figure}

\begin{figure}
 \includegraphics[width=15cm,angle=270]{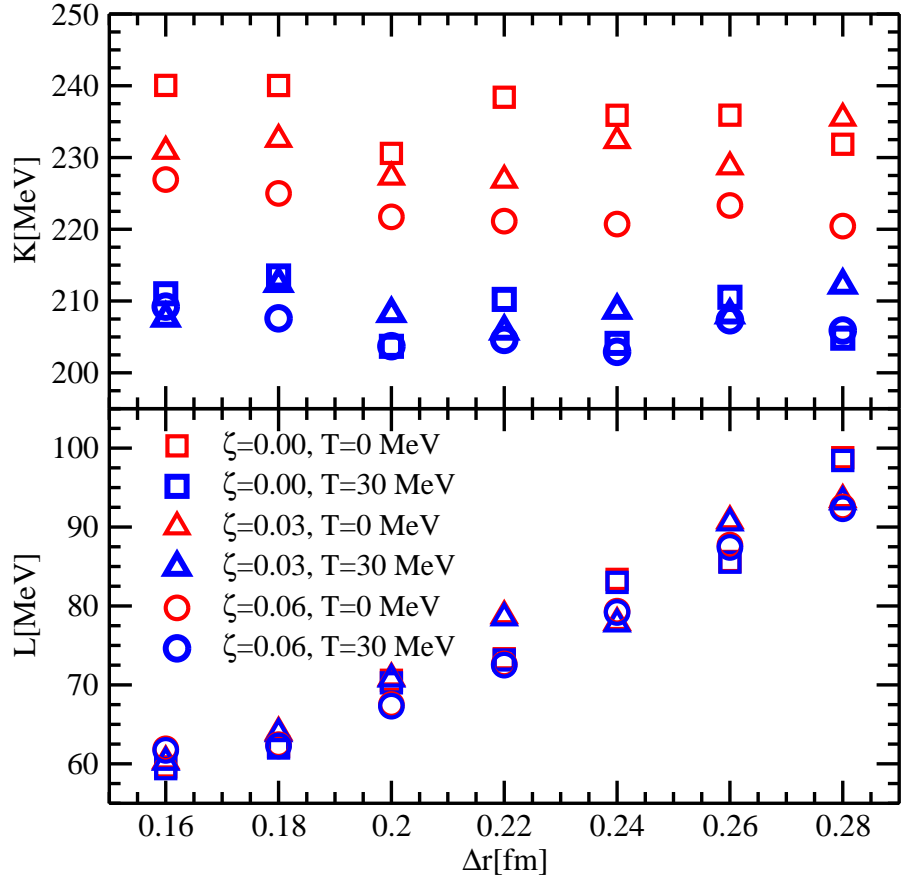}
  \caption{\label{fig:fig2}
(Color online) Same as Fig. 1, but for the slope of the symmetry
energy (L) and incompressibility coefficient (K) of nuclear
matter.}
\end{figure}

The nuclear symmetry energy is a fundamental input to understand
the exotic nuclei, heavy ion collision data and many other
astrophysical phenomena. Therefore recently many efforts have been
made to extract the information on the magnitude and density
dependence of symmetry energy of nuclear matter. In Fig.
\ref{fig:fig1} we present the values of $E_{sym}(\rho_0)$ at
saturation density as a function of $\Delta$r, the neutron skin
thickness in the $^{208}Pb$ nucleus for various model
parametrizations, the square represent the parametrizations
BSR1-BSR7 with $\zeta$ = 0.00, the triangles represent the
parametrizations BSR8-BSR14 with $\zeta$ = 0.03, and the circles
represent the parametrizations BSR15-BSR21 with $\zeta$ = 0.06. In
Fig. \ref{fig:fig2}, in the lower panel we present the slope of
symmetry energy and in the upper panel we present the
incompressibility coefficient  for nuclear matter as a function of
$\Delta$r. In Figs. \ref{fig:fig1} and \ref{fig:fig2} the red
symbols represent the results at T = 0 MeV and the blue symbols
represent the results at T = 30 MeV. It is found that variation in
the values of symmetry energy becomes reasonably large as the
value of neutron skin thickness increases, where as the value for
the slope of symmetry energy remains unaffected at temperature T =
0 and 30 MeV. The value of incompressibility coefficient is
sensitive to $\zeta$ and indicates the change at T = 30 MeV.

\begin{figure}
 \includegraphics[width=15cm,angle=270]{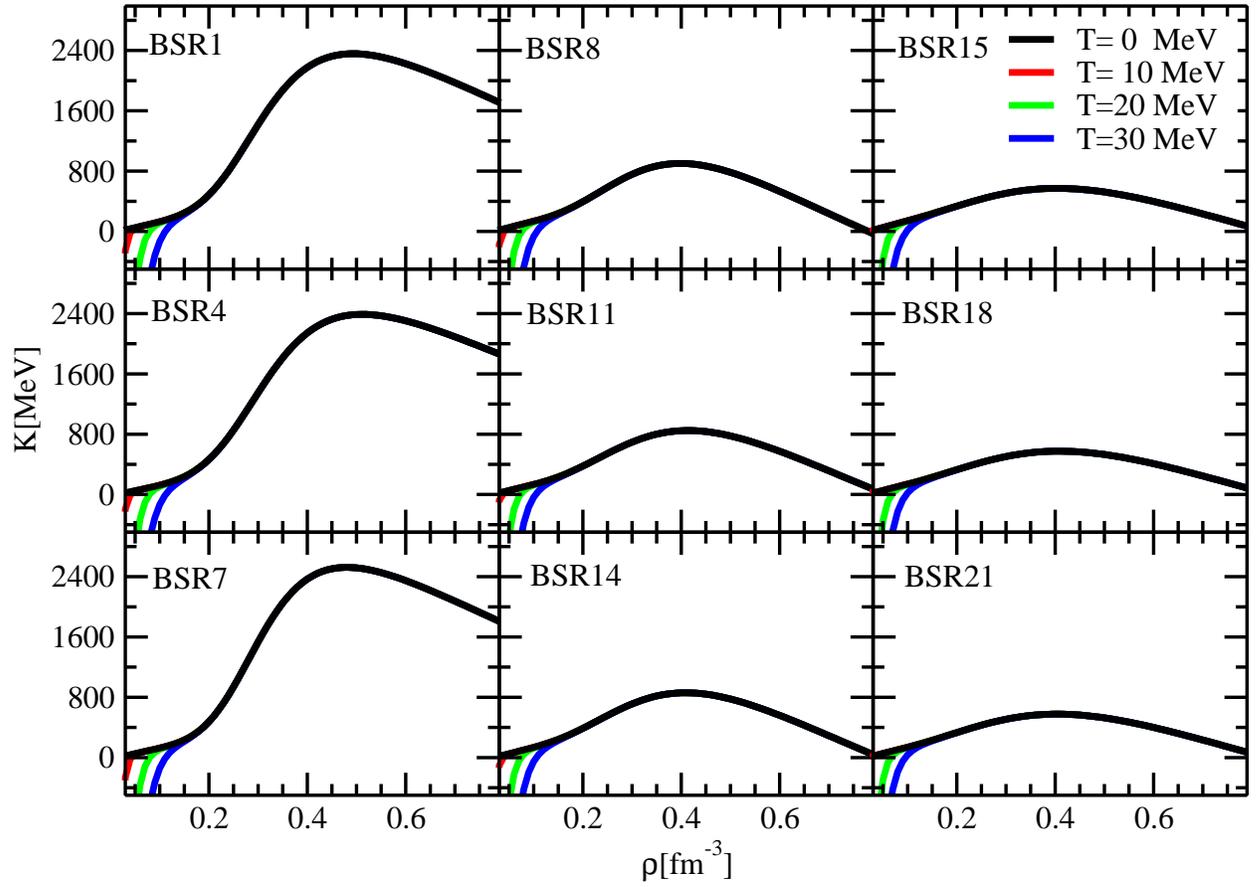}
  \caption{\label{fig:fig3}
(Color online) The density dependence of the incompressibility
coefficient is plotted at temperatures of 0, 10, 20, and 30 MeV
for various parametrizations.}
\end{figure}

\begin{figure}
 \includegraphics[width=15cm,angle=270]{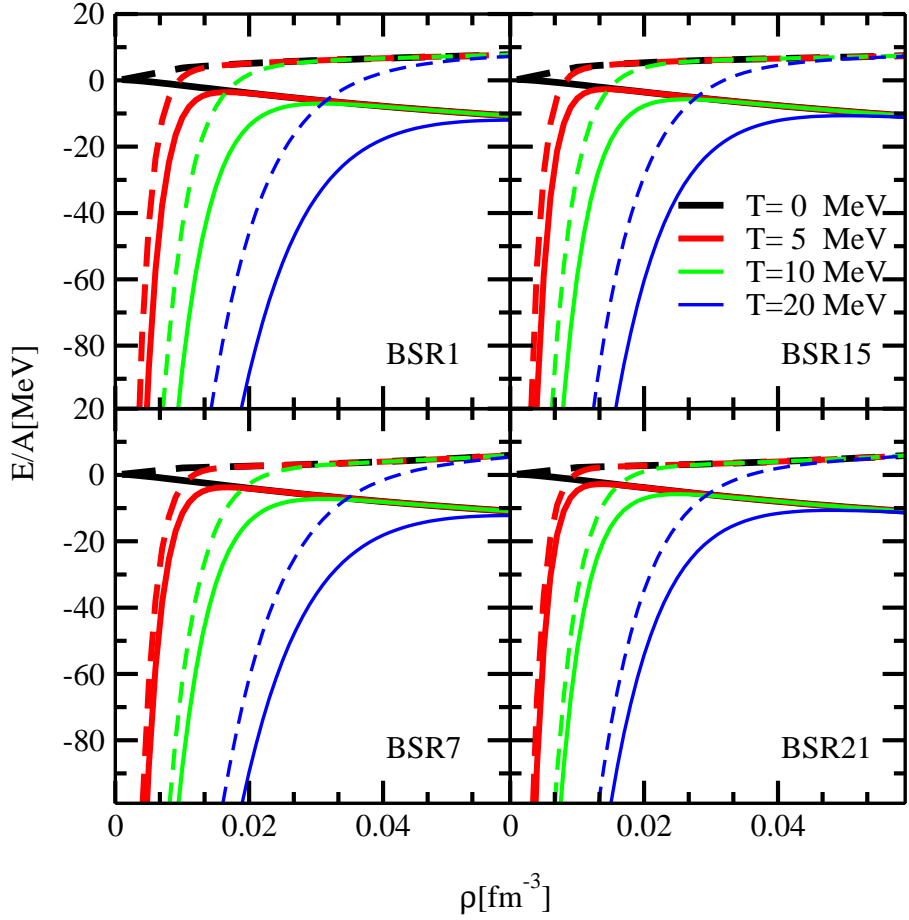}
  \caption{\label{fig:fig4}
(Color online) The variation in energy per nucleon (E/A) for
symmetric nuclear matter (solid lines) and for pure neutron matter
(dashed lines) calculated with the BSR1, BSR7, BSR15, and BSR21
parametrizations is plotted as a function of density at
temperatures of 0, 5, 10, and 20 MeV.}
\end{figure}

\begin{figure}
 \includegraphics[width=15cm,angle=270]{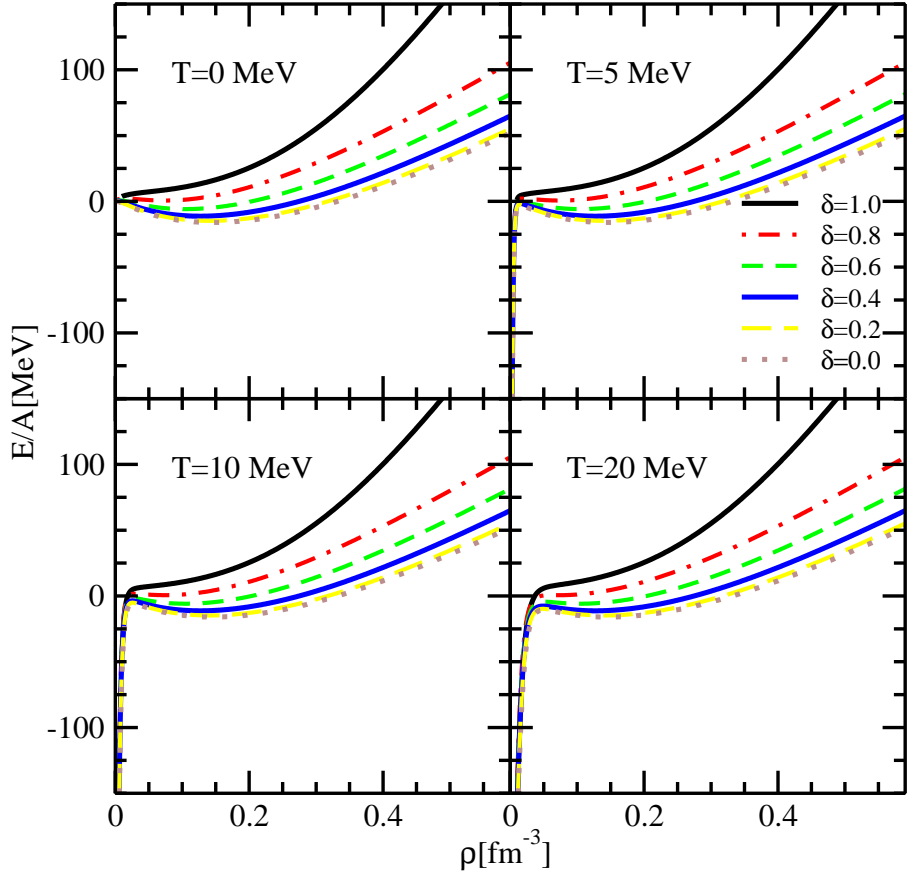}
  \caption{\label{fig:fig5}
(Color online) The variation in energy per nucleon (E/A) for
asymmetric nuclear matter at temperatures of 0, 5,10, and 20 MeV
with various values of $\delta$ calculated with the BSR15
parametrization.}
   \end{figure}

In Fig. \ref{fig:fig3} we compare the density dependence of the
incompressibility coefficient at finite temperatures for various
parametrizations with cold nuclear matter. It is found that the
incompressibility coefficient at finite temperature has shown
change below neutron saturation densities $\rho_0$ = 0.15
fm$^{-3}$ only, and K gain maximum value in the range of densities
of  $\sim 0.4$ to $0.5 fm^{-3}$. The maximum value is very
sensitive to $\zeta$, remains almost same on varying $\Delta$r,
and decreases with increasing temperature. Further, we explore the
effect of density on energy per nucleon (E/A) for symmetric
nuclear matter and pure neutron matter at finite temperatures as
shown in Fig. \ref{fig:fig4}, computed by employing the BSR1,
BSR7, BSR15, and BSR21 parametrizations. At the finite temperature
the E/A for symmetric nuclear matter decreases sharply as compared
to the E/A for pure neutron matter in the low density regime, and
with the increase of $\zeta$ from 0.00 to 0.06, the E/A of both
symmetric nuclear matter and pure neutron matter decreases
moderately. The value of E/A remains almost unchanged for the
variation in values of $\Delta$r in $^{208}$Pb. In Fig.
\ref{fig:fig5} we present the variation of E/A as a function of
density at the different values of the asymmetry parameter
$\delta$ at T = 0, 5, 10, and 20 MeV for the BSR15
parametrization. The value of E/A increases reasonably well with
the increase in value of $\delta$. Fig. \ref{fig:fig5} shows that
the E/A changes in the higher density region due to the change in
the asymmetry parameter $\delta$, whereas in the low density
region E/A varies with increases in temperature.

\begin{figure}
 \includegraphics[width=15cm,angle=270]{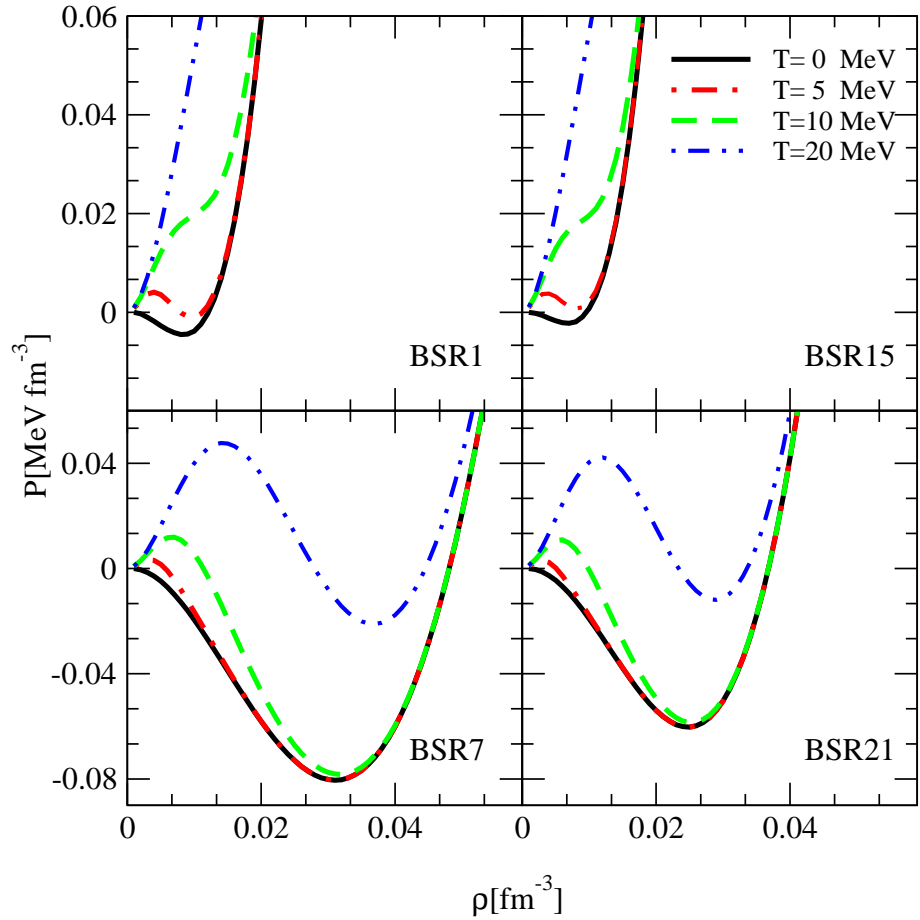}
  \caption{\label{fig:fig6}
(Color online) The pressure for symmetric nuclear matter for the
BSR1, BSR7, BSR15, and BSR21 parametrizations is plotted as a
function of density at temperatures of 0, 5, 10, and 20 MeV.}
   \end{figure}

\begin{figure}
 \includegraphics[width=15cm,angle=270]{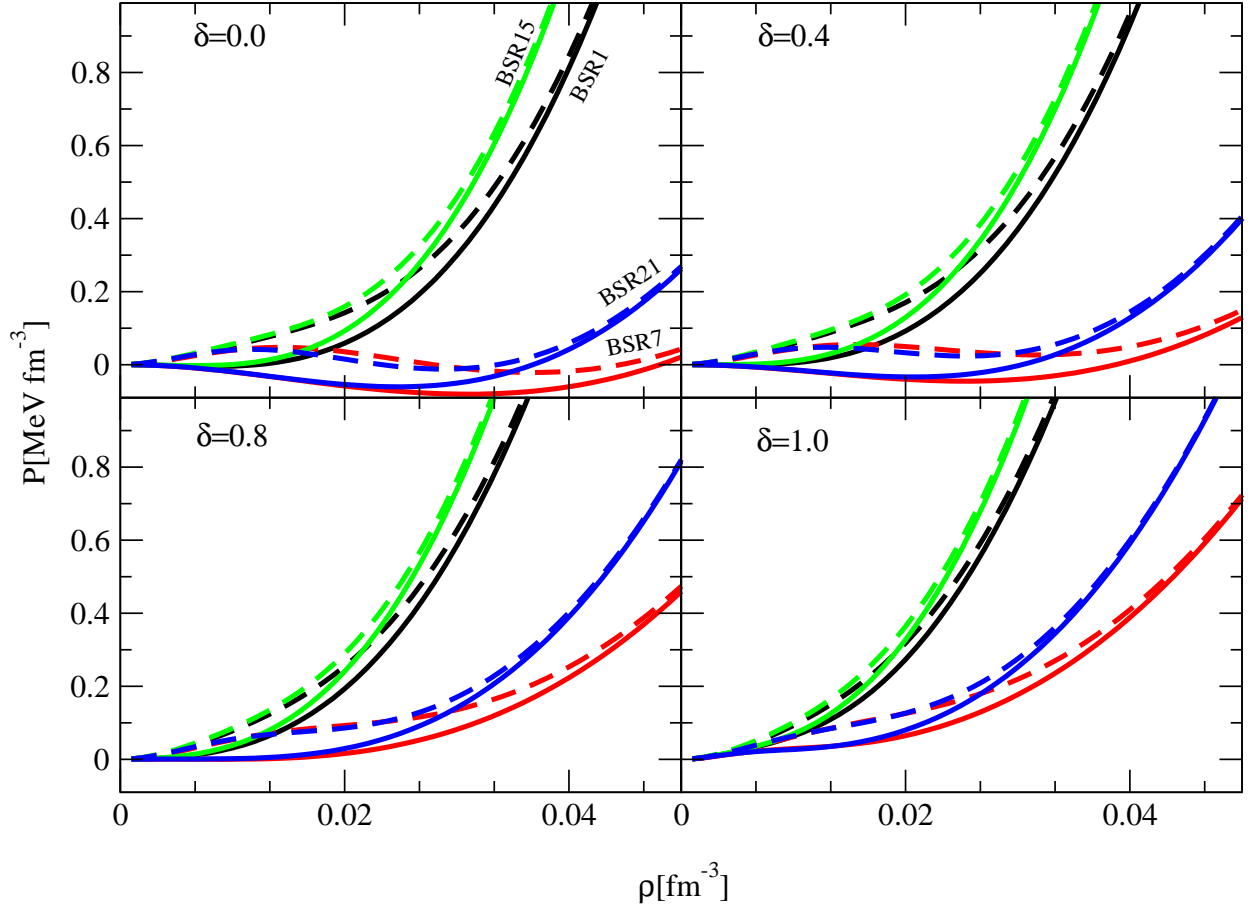}
  \caption{\label{fig:fig7}
(Color online) The pressure of asymmetric nuclear matter plotted
as a function of density in low density regions for various values
of the asymmetry parameter $\delta$. The solid line represents T =
0 MeV and the dashed line represents T = 20 MeV. The black line,
red line, green line, and blue line represent the BSR1, BSR7,
BSR15, and BSR21 parametrizations respectively.}
   \end{figure}

In Fig. \ref{fig:fig6} we present the variation of the equation of
state of symmetric nuclear matter as a function of nuclear matter
density at various temperatures for the BSR1, BSR7, BSR15, and
BSR21 parametrization, in the very low density region. The
pressure varies with temperatures at small values of densities and
has negligible effect at higher densities. The variation in
pressure for a given density depends mostly on the choice of
parametrizations and temperature. The pressure become negative for
BSR7 and BSR21 parametrizations with $\Delta$r = 0.28fm, in the
low density regime ($\rho \le 0.04 fm^{-3}$). In Fig.
\ref{fig:fig7} the pressure of asymmetric nuclear matter is
plotted as a function of density in the low density region for
various values of the asymmetry parameter $\delta$. The solid line
represents T = 0 MeV and the dashed line represent T = 20 MeV. The
black line, red line, green line, and blue line represent the
BSR1, BSR7, BSR15, and BSR21 parametrization respectively. The EOS
becomes stiff with the increase in the asymmetry parameter
$\delta$ and trend continues till it becomes pure neutron matter.

\begin{figure}
 \includegraphics[width=15cm,angle=270]{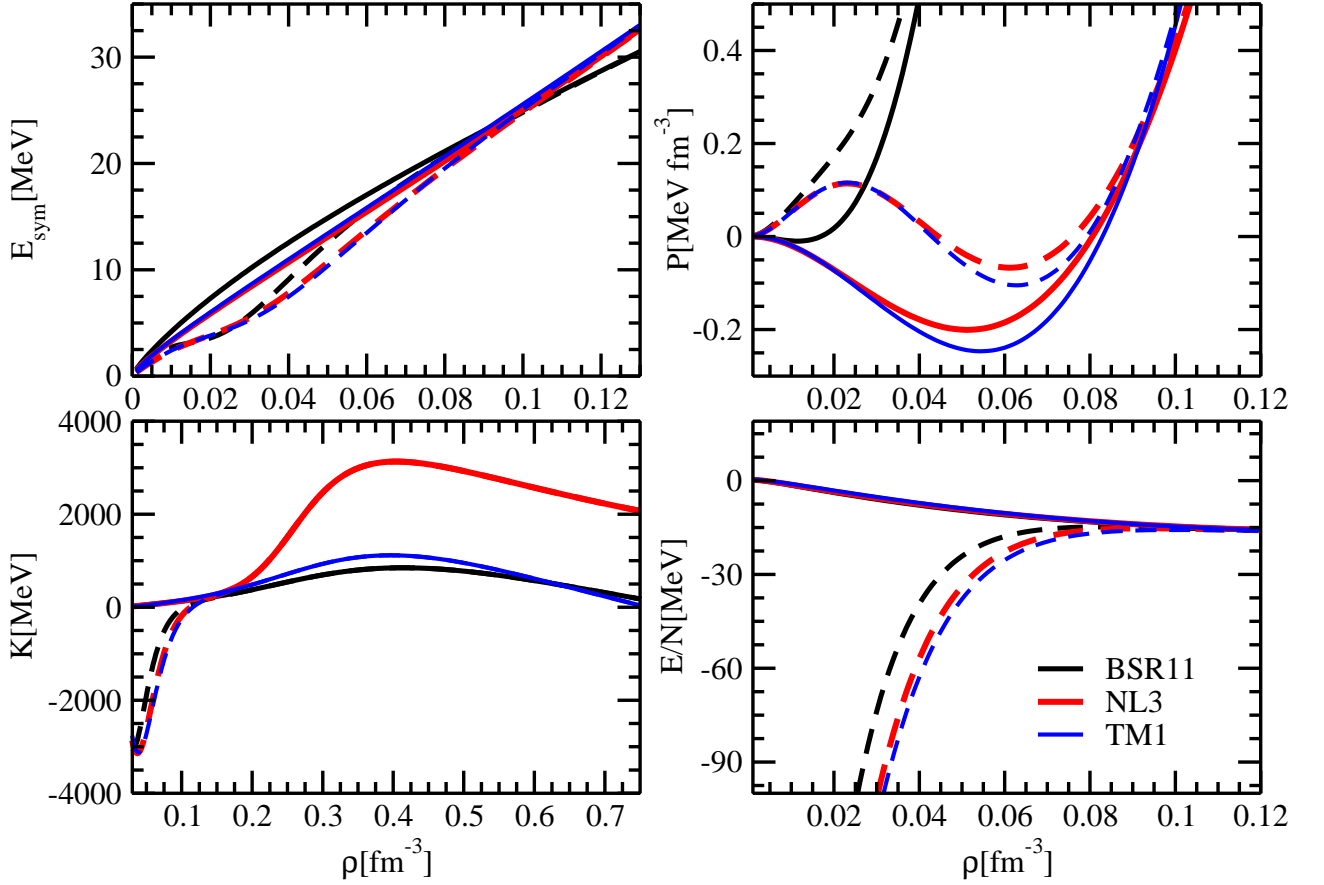}
  \caption{\label{fig:fig8}
(Color online) The comparison of the theoretical results for
symmetry energy ($E_{sym}$), the incompressibility coefficient
(K), pressure (P) and energy per nucleon (E/A) computed with the
BSR11,
 NL3, \cite{Lalazissis97} and TM1 \cite{Sugahara94} parametrizations
of relativistic mean field theory at temperatures of 0 and 30 MeV
as a function of density. The black lines represent the results of
the BSR11 parametrization,  the red lines represent the NL3
parametrization, and the blue lines represent the TM1
parametrization. The solid lines and the dashed lines represent
temperatures of 0 and 30 MeV, respectively. }
   \end{figure}

We study the density dependence of symmetry energy, nuclear matter
pressure density, and energy per nucleon at low density and the
nuclear matter incompressibility coefficient as a function of
density, with different RMF models. The comparison of the
theoretical results for $E_{sym}$, P, E/A and K computed with
BSR11 with  NL3\cite{Lalazissis97} and TM1\cite{Sugahara94}
parametrizations of RMF theory at temperatures of 0 and 30 MeV as
a function of density are presented in Fig. \ref{fig:fig8}. The
solid and dashed black lines represent the results of the BSR11
parametrization,  the red lines represent the NL3 parametrization,
and the blue lines represent the TM1 parametrization. The solid
lines and dashed lines represent temperatures of 0 MeV and 30 MeV,
respectively. It is found that the values of $E_{sym}$, P, E/A and
K are very sensitive to temperatures at lower densities ($\sim
0-0.1 fm^{-3}$) and,  are independent of temperature at higher
densities. Further, we find a reasonable  change in the behavior
of symmetry energy for small values of $\Delta$r. However, the
symmetry energy decreases at $30$ MeV, and at very low density
($\sim 0.02 fm^{-3}$) for the value of $E_{sym}$ the trend
reverses as shown in Fig. \ref{fig:fig8}. It is noteworthy from
Fig. \ref{fig:fig8} that, except for nuclear matter
incompressibility computed with the NL3 parametrization, all other
RMF parametrizations yield almost the same values of bulk
properties.

\subsection{Non Rotating PNS}

We discuss the properties of protoneutron star composed of charge
neutral nuclear matter at different temperatures. The fixed total
baryon density is given as,
\begin{equation}
\label{eq:tden}
\rho=\sum_B \rho_B,
\end{equation}
the charge neutrality condition is given as
\begin{equation}
\label{eq:chnut}
\sum_Bq_B\rho_B+\sum_Lq_L\rho_L=0,
\end{equation}
and the chemical equilibrium conditions,
\begin{equation}
\label{eq:chemnut}
 \mu_B=\mu_N-q_B\mu_e,
\end{equation}
\begin{equation}
\label{eq:chemele}
\mu_{\mu}=\mu_e
\end{equation}
are satisfied. For density higher than $0.5\rho_0$ the baryonic
part of the EOS  is evaluated within the ERMF model, whereas the
contributions of the electrons and muons to the EOS are evaluated
within the Fermi gas approximation. At densities lower than
$0.5\rho_0$ down to $0.4$ x $10^{-10} \rho_0$ we use the EOS of
Baym et al. \cite{Baym71}. The properties of nonrotating compact
star are obtained by integrating the Tolman-Oppenheimer-Volkoff
equations \cite{Weinberg72}.

\begin{figure}
\includegraphics[width=15cm,angle=270]{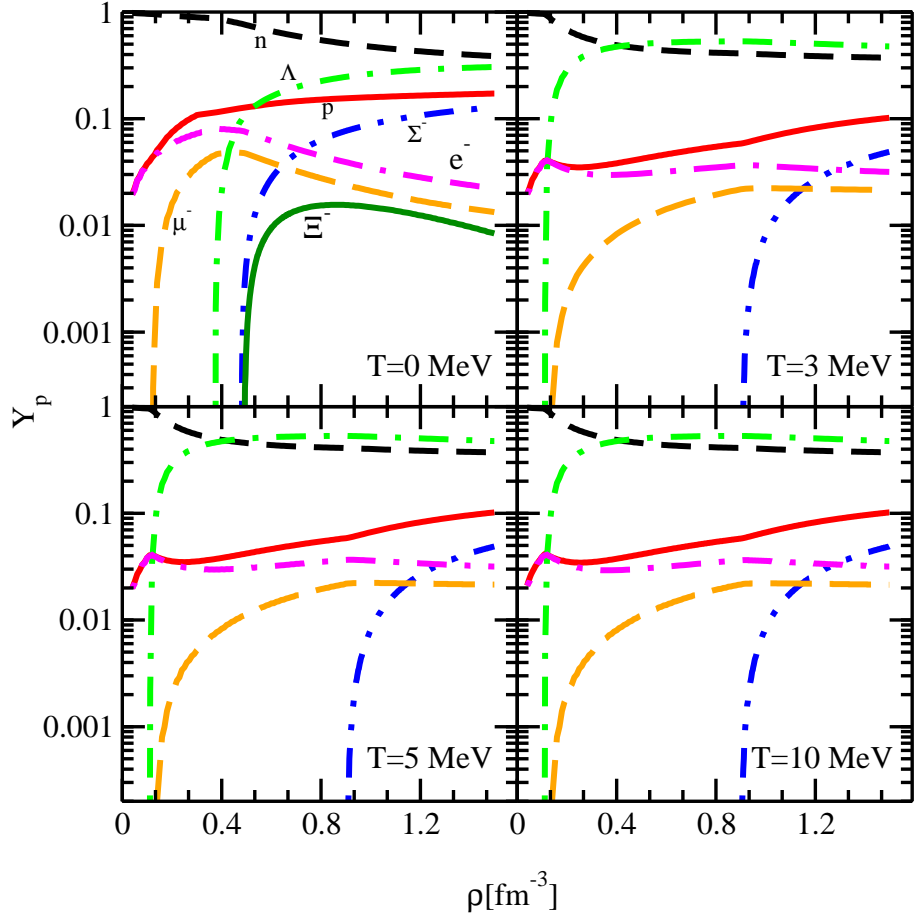}
\caption{\label{fig:fig9} (Color online) Relative particle
fraction as a function of baryon density of the compact stars
obtained for the BSR15 parametrization with the hyperon meson
coupling parameter $X_{\omega y}$ = 0.50 at different
temperatures.}
\end{figure}

\begin{figure}
 \includegraphics[width=15cm,angle=270]{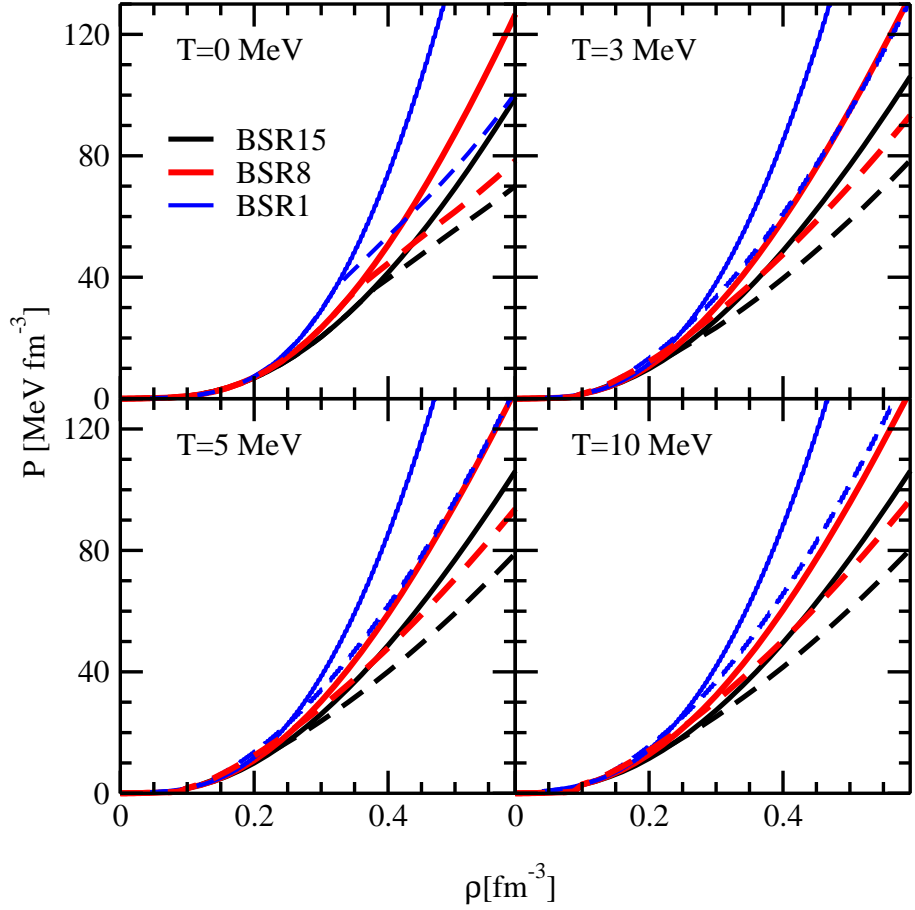}
  \caption{\label{fig:fig10}
(Color online) The pressure density is plotted as a function of
density employing using the BSR1,BSR8, and BSR15 parametrizations.
The dashed lines represent EOS with hyperons having the
hyperon-meson coupling parameter $X_{\omega y}$ = 0.50 and solid
lines represent EOS without hyperons at temperatures of 0, 3, 5,
and 10 MeV.}
 \end{figure}

Fig.\ref{fig:fig9} shows the relative particle fraction calculated
at different temperatures for the BSR15 parametrization as a
function of density. At finite temperature, neutrons, protons,
$\Lambda$ hyperons, and electrons become abundant at baryon
density lower than their particle threshold density in the cold
nuclear matter, where as $\Xi$ hyperons disappears even at T = 3
MeV and the particle threshold densities of muons and $\Sigma$
hyperons increase to $0.902$ and $0.5fm^{-3}$, respectively, as
compared with their threshold densities in cold nuclear matter. In
our calculation, the threshold densities of hyperons in cold
matter are as follows: for $\Lambda$-hyperons the threshold
density is 0.376 $fm^{-3}$, for $\Sigma^-$ hyperons it is 0.482
$fm^{-3}$ and for $\Xi^-$ hyperons it is 0.490 $fm^{-3}$, but at T
= 10 MeV the threshold density of $\Lambda$ hyperons decreases to
$0.112fm^{-3}$, of $\Sigma^-$ hyperons it increases to
$0.902fm^{-3}$ and for  $\Xi^-$ it disappears as shown in Fig.
\ref{fig:fig9}. We also observed the effect of temperature on
relative particle fraction in compact stars without hyperons. It
is found that with the increase in temperature the neutrons,
protons and leptons become abundant at lower densities however, at
higher densities for protons, electrons, and muons, the magnitude
of the particle fraction slightly decreases. The pressure density
is plotted as a function of baryon density by employing BSR1, BSR8
and BSR15 parametrizations as shown in Fig. \ref{fig:fig10}. The
dashed lines represent EOS with hyperons having hyperon-meson
coupling parameter $X_{\omega y}$ = 0.50 and the solid lines
represent EOS without hyperons at temperatures of T = 0, 3, 5, and
10 MeV. The EOS become stiff at higher temperature with and
without inclusion of hyperons and, subsequently, there is an
increase in the gravitational mass of the CS.

\begin{figure}
\includegraphics[width=15cm,angle=270]{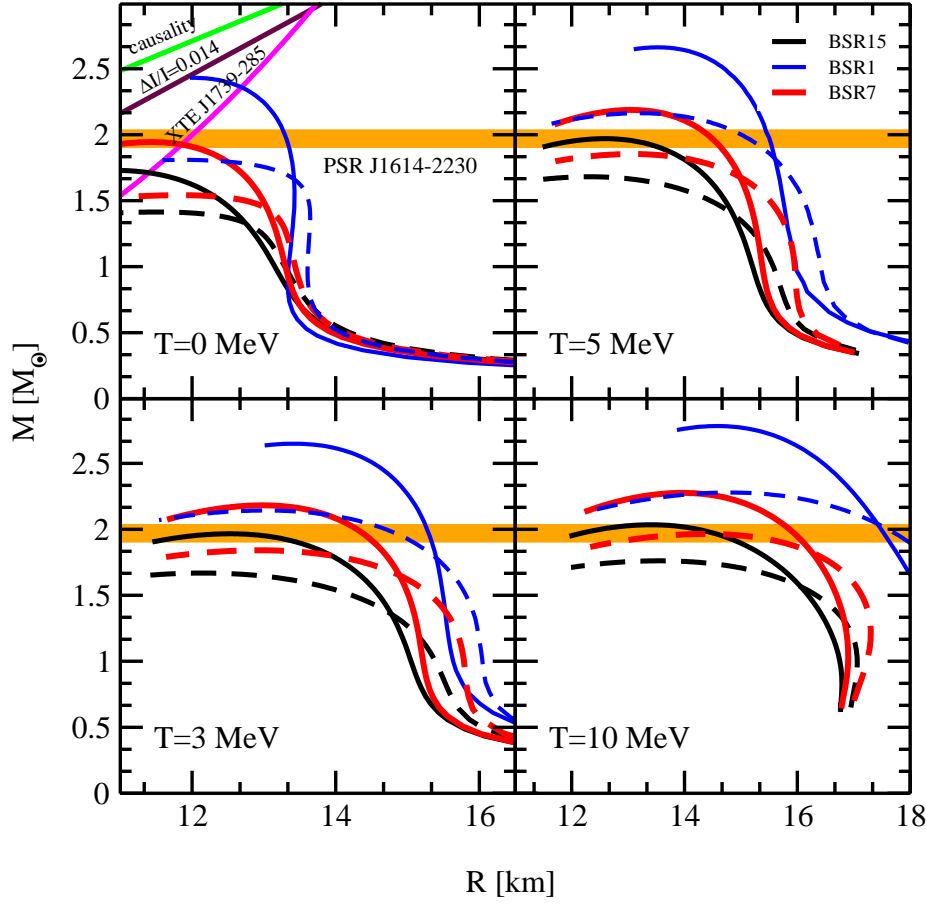}
\caption{\label{fig:fig11} (Color online) The mass and radius
relationship of protoneutron stars. The dashed lines represent
mass for EOS with hyperons having the hyperon-meson coupling
parameter $X_{\omega y}$ = 0.50, and the solid lines represent
mass for EOS without hyperons at temperatures of 0, 3, 5, and 10
MeV. The colors blue, red, and black represent the BSR1, BSR8 and
BSR15 parametrizations respectively. The recent mass measurement
of the PSR J1614-2230 pulsar is also displayed.}
\end{figure}

In Fig. \ref{fig:fig11} we present the gravitational mass and
radius relationship for the PNS. The dashed lines represent mass
for EOS with hyperons at $X_{\omega y}$ = 0.50 and the solid lines
represent mass for EOS without hyperons at temperatures of 0, 3,
5, and 10 MeV. The region excluded by causality (green solid line)
and rotation constrains of compact star XTE J1739-285 (maroon
solid line) are shown in the upper left panel. The mass and radius
limit estimate from Vela pulsar glitches $\Delta$I/I=0.014 is
shown as the magenta solid line in the upper left panel. the
recent mass measurement of  the PSR J1614-2230 pulsar of
1.97$\pm$0.04 \cite{Demorest10} is displayed in Fig.
\ref{fig:fig11} as another constrain to the nuclear matter EOS,
computed by our group in Ref.\cite{Dhiman07}. The EOS that contain
exotic hadronic matter of hyperons does not satisfy the mass
constrain of the PSR J1614-2230 as shown in Fig. \ref{fig:fig11},
and also as discussed similar in Ref. \ref{Demorest10} that the
model of EOS includes the appearance of hyperons or kaon
condensates.
 However the EOS computed\cite{Dhiman07} without hyperons for
 $\zeta$ = 0.00 and $\zeta$ = 0.03 satisfy the constraint of the PSR J1614-2230 pulsar mass measurement
and the prediction of its radius 11-15 km, where $\zeta$ is the
$\omega$-meson self coupling parameter, and mainly affects the
high density behavior of the EOS and cannot be constrained by the
structural properties of finite nuclei measurements and bulk
properties of nuclear matter at saturation density. From the
argument of Tolman VII solution of Einstein's equations for the
relationship between maximum gravitational mass and its upper
limit on the central energy density, we get the value
\cite{Dhiman10} of $\epsilon_c$=1.92 $\times$10$^{15}$ gcm$^{-3}$
and 2.73$\times$10$^{15}$gcm$^{-3}$ for BSR1 and BSR15
parametrization respectively. The EOSs of warm dense nuclear
matter becomes stiffer than the EOS of cold dense nuclear matter
of the compact stars, even in the presence of exotic matter.

\begin{figure}
 \includegraphics[width=13cm,angle=270]{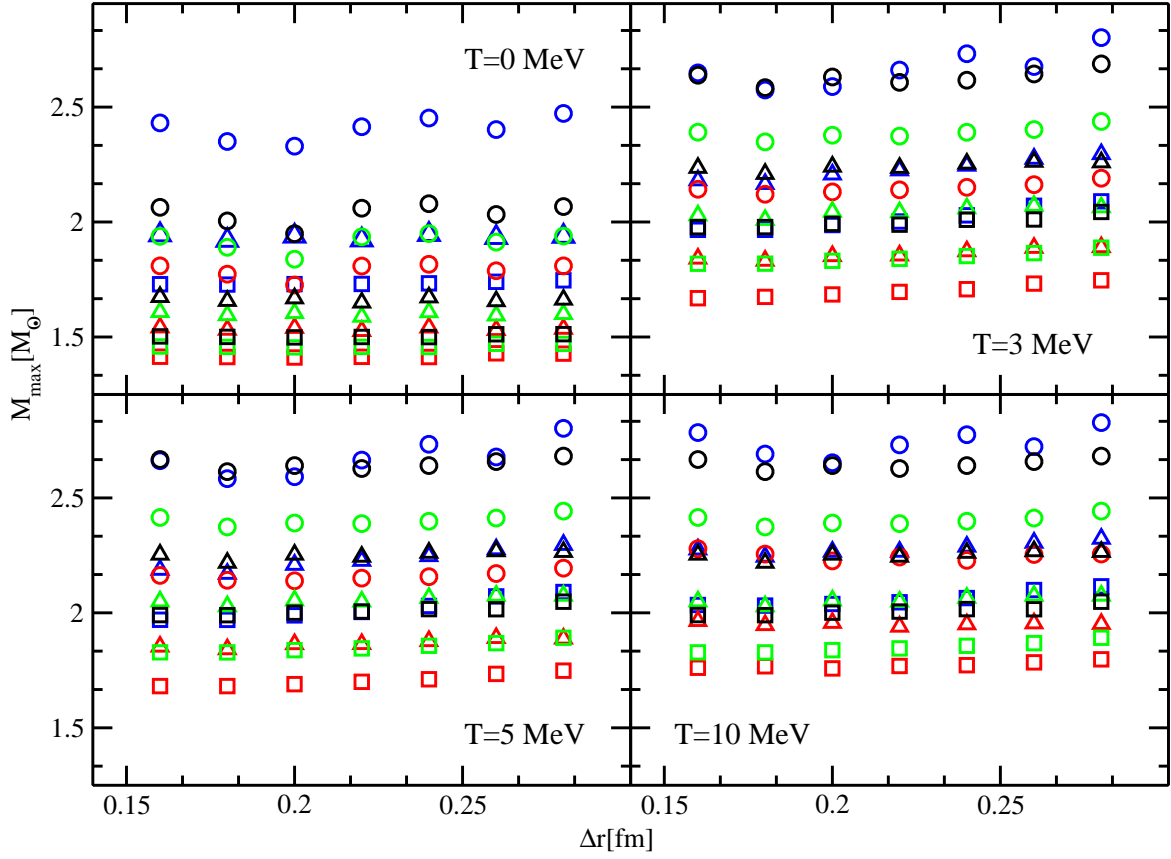}
  \caption{\label{fig:fig12}
(Color online) The maximum gravitational mass of protoneutron star
is plotted as function of neutron skin thickness ($\Delta$r) in
the $^{208}Pb$ nucleus at temperatures of 0, 3, 5, and 10 MeV. The
color blue represents the masses of EOS without hyperons whereas
the color red represents  masses of EOS having hyperons with
hyperon-meson coupling parameter $X_{\omega y}$ = 0.50 . The
colors green and black represent  masses of EOS having hyperons
with $X_{\omega y}$ 0.60 and 0.70, respectively. The circles,
triangles, and squares represent the values of maximum
gravitational mass for the $\omega$-meson self-coupling $\zeta$ =
0.0, 0.03, and 0.06, respectively.}
\end{figure}

In Fig.\ref{fig:fig12} the maximum gravitational mass of
protoneutron star is plotted as a function of neutron skin
thickness $\Delta$r in the $^{208}Pb$ nucleus at temperature of 0,
3, 5, and 10 MeV. The color blue represents the mass computed with
EOS without hyperons whereas red represents gravitational masses
of EOS including hyperon, at $X_{\omega y}$ = 0.50. Green and
black represent  masses of EOS with hyperon, the values of
$X_{\omega y}$ are equal to 0.60 and 0.70, respectively. The
circles, triangles, and squares represent the values of maximum
gravitational mass for the $\omega$-meson self-coupling $\zeta$ =
0.0, 0.03 and 0.06, respectively. We varied the hyperon meson
coupling parameter $X_{\omega y}$ from 0.50 to 0.70 at all
temperatures and found that on increasing the coupling parameter
the maximum gravitational mass of the PNS increased. It is noticed
that the increase in gravitational mass of the compact star is
large at $\zeta$=0.00 and reasonably small at $\zeta$=0.06.
Further, Fig.\ref{fig:fig12} shows that the cold compact star with
hyperons can have gravitational mass $M \ge 2 M_{\odot}$ if
$\zeta$ = 0.00 and $X_{\omega y} \ge$ 0.70, whereas the
protoneutron star with hyperons can have mass $M \ge 2 M_{\odot}$
if the chosen parameters are $\zeta$ = 0.00 or 0.03 and $X_{\omega
y}\ge$  0.50, and the compact star satisfies the constraint of the
mass measurement of the PSR J1614-2230 pulsar \cite{Demorest10}.

In Tables \ref{tab2}-\ref{tab4} we have presented the key
structural properties of compact stars at finite temperature;
properties such as maximum gravitational mass, radius at maximum
gravitational mass, radius for star with canonical mass ($1.4
M_\odot$), and gravitational redshift of the photon $Z_{surf}$
emitted from the compact star surface for a star with maximum mass
and canonical mass using BSR1-BSR21 parametrizations at
temperatures of 0, 3, 5, and 10 MeV without and with the inclusion
of hyperons for $X_{\omega y}$=0.50 only. It is observed from the
Tables \ref{tab2}-\ref{tab4} that with an increase in the value of
the $\zeta$ parameter the mass of the compact star decreases,
whereas with a rise in temperature the mass of the compact star
increases. Further with increasing $\Delta$r the mass increases
for all temperatures. It is also observed that when the
temperature changes from 0 to 3 MeV there is an increase in the
mass of the compact star by $\sim 0.2$ to $0.4 M_\odot$, but on
further increasing temperature T $\ge$ 5 MeV this increase in mass
of compact star becomes very small. Also the increase in the
radius at maximum mass and canonical mass ($M_{max}$ and
$M_{1.4}$) with temperature is $\sim 1.5-2$ km initially but
becomes smaller with further increase in temperature. The radius
also increased on increasing $\Delta$r for all temperatures but
decreased with increase in the value of the $\zeta$ parameter. The
results for gravitational redshift of the photon Z$_{surf}$
emitted from the surface of the compact star, can be computed as;
\begin{equation}
\label{zsurf}
Z_{surf} = \left(1- \frac{2GM}{Rc^2}\right)^{-1/2} -1,
\end{equation}
where R is the radius and M is the gravitational mass of the
compact star. It is clear from Tables \ref{tab2}-\ref{tab4} that,
with an increase in the $\zeta$ or $\Delta$r parameter, the
gravitational redshift $Z_{surf}$ and $Z_{1.4}$ decrease, whereas
with an increase in temperature $Z_{surf}$ and $Z_{1.4}$ decrease
further. It is observed that the inclusion of hyperons in PNS
decreases the magnitude of the mass, the radius and the
gravitational redshift for all RMF parametrizations. In comparison
with cold compact star, we obtained that the gravitational mass of
the PNS with and without hyperons increases by  $\sim
0.4M_{\odot}$, its radius increases by $\sim 3$km, and the radius
$R_{1.4}$ at the canonical mass of the computed data increases by
3-6km, where as the value of the gravitational red shift  at
finite temperature decreases approximately 0.03-0.07.

\subsection{Rotating PNS}
\begin{figure}
\includegraphics[width=13cm,angle=270]{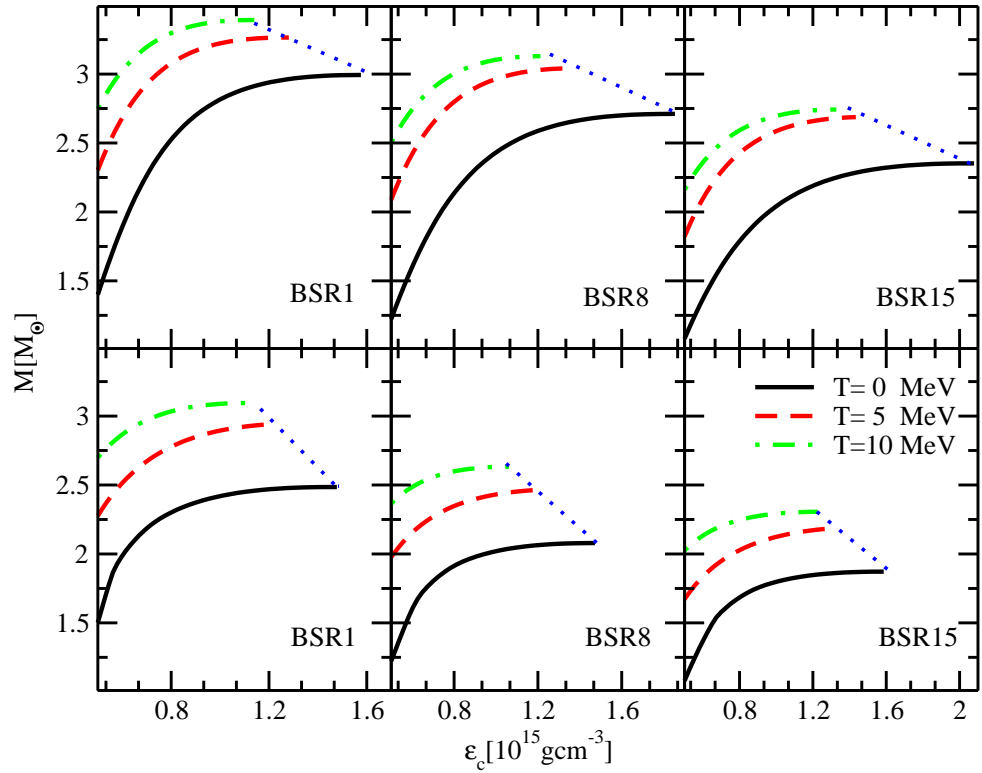}
\caption{\label{fig:fig13} (Color online) The mass shedding limit
(Kepler) is plotted for EOS obtained using  BSR1, BSR8, and BSR15
parametrizations at temperatures of 0, 5, and  10 MeV in terms of
gravitational mass M as a function of central energy density
$\epsilon_c$. The upper panels contain EOS without hyperons,
whereas the lower panels contain EOS with hyperons having the
hyperon-meson coupling parameter $X_{\omega y}$ = 0.50. The
slanting dotted (blue) line corresponds to the axisymmetric
instability limit.}
\end{figure}

The keplerian configurations of rapidly rotating PNS have been
computed in the framework of general relativity by solving the
Einstein field equations for stationary axisymmetric space time (
e.g., see Ref.\cite{Stergioulas03} and references therein). The
numerical calculations have been performed by employing the
rotating neutron star (RNS) code \cite{Stergioulas95}. In
Fig.\ref{fig:fig13} the mass shedding limit (Kepler) is plotted
for EOS obtained by using  the BSR1, BSR8, and BSR15
parametrizations at 0, 5, and  10 MeV  in terms of gravitational
mass M as a function of central energy density $\epsilon_c$. The
upper panel contains EOS without hyperons, whereas the lower panel
contain EOS with hyperons at $X_{\omega y}$ = 0.50. Keplerian
configurations terminate at the central energy density where
equilibrium solutions are stable with respect to the small
axisymmetric perturbations; the slanting dotted (blue) line
corresponds to the axisymmetric instability limit. In the Kepler
limit sequences, the gravitational maximum mass of the PNS
increase with increases in temperature by $20\%-23\%$ and its
corresponding equitorial radius increases by $ 25\%-46\%$, with
respect to its non rotating gravitational maximum mass and radius,
respectively. These observations are reasonably well within the
predictions provided in Refs. \cite{Stergioulas03,Goussard97} and
are slightly higher in the case of the PNS with hyperons. Compared
with the cold nuclear matter compact star, the Keplerian angular
velocity of the PNS decreases by $5\%-8\%$ in the case of the PNS
without hyperons, and it is $14\%-20\%$ for the PNS with hyperons.

\section{Conclusion}

The effect of temperature and density dependence of  the
asymmetric nuclear matter properties is studied within the ERMF
model which includes the contribution from the self and mixed
interaction terms by using different parametrizations obtained by
varying the neutron skin thickness $\Delta$r and the
$\omega$-meson self coupling ($\zeta$). We studied the bulk
properties of cold and warm nuclear dense matter at finite
temperature, and compared the structural properties of non
rotating and rotating cold compact stars with PNS, constructed
within ERMF model.

We observed that the changes in bulk properties at saturation
densities are negligible till a temperature of 20 MeV but as
temperatures increase further these properties start varying
significantly. It is found that variation in the values of
symmetry energy becomes reasonably large as the value of the
neutron skin thickness increases, where as the value for the slope
of symmetry energy remains unaffected at T = 0 and 30 MeV. The
value of the incompressibility coefficient is sensitive to $\zeta$
and indicates the change at T = 30 MeV. the energy per nucleon for
symmetric nuclear matter decreases sharply as compared to the
energy for pure neutron matter at very low densities, and upon
increasing the $\zeta$ the decrease becomes moderate whereas upon
increasing $\Delta$r the value remains almost the same. It is
observed that with the increase in the asymmetry parameter
$\delta$, the EOS become stiff and the trend continues until it
becomes pure neutron matter.
 It is found that the
temperature dependence of the symmetry energy is more sensitive to
the small values of $\Delta$r. Although the symmetry energy
decreases with increases in temperature, at very low density
($\sim 0.02 fm^{-3}$) the trend reverses.

In our calculations at finite temperature, neutrons, protons,
$\Lambda$ hyperons, and electrons become abundant at baryon
density lower than their particle threshold density in the cold
nuclear matter, where as $\Xi$ hyperons disappears even at T = 3
MeV.  The EOS of warm dense nuclear matter becomes stiffer than
the EOS of the cold dense nuclear matter of the compact stars,
even if we include the exotic matter.  We varied the hyperon meson
coupling parameter $X_{\omega y}$ from 0.50 to 0.70 at all
temperatures and found that on increasing the coupling parameter
the maximum gravitational mass of the star increased and this
increase was large for the smaller values of $\zeta$ and small for
the larger values of $\zeta$.  Values of all the properties such
as mass, radius, and gravitational redshift decreased upon
inclusion of hyperons for all the parametrizations. We obtained
that the gravitational mass of the PNS with and without hyperons
increased by $\sim 0.4M_{\odot}$ and its radius increased by $\sim
3$km, and the radius $R_{1.4}$ at the canonical mass of the
computed data increased by 3-6 km, where as the value of the
gravitational red shift  at finite temperature decreased
approximately 0.03-0.07. In the Kepler limit sequences, the
gravitational maximum mass of PNS increase with increase in
temperature by $20\%-23\%$ and its corresponding equitorial radius
increased by $ 25\%-46\%$, with respect to its non rotating
gravitational maximum mass and radius, respectively. The Keplerian
angular velocity of PNS without hyperons decreased by $5\%$- $8\%$
and it decreased by $14\%$-$20\%$ for PNS with hyperons, in
comparison to the cold CS without and with hyperons, respectively.

\newpage
\bibliography{1review}

\begin{thebibliography}{35}
\expandafter\ifx\csname natexlab\endcsname\relax\def\natexlab#1{#1}\fi
\expandafter\ifx\csname bibnamefont\endcsname\relax
  \def\bibnamefont#1{#1}\fi
\expandafter\ifx\csname bibfnamefont\endcsname\relax
  \def\bibfnamefont#1{#1}\fi
\expandafter\ifx\csname citenamefont\endcsname\relax
  \def\citenamefont#1{#1}\fi
\expandafter\ifx\csname url\endcsname\relax
  \def\url#1{\texttt{#1}}\fi
\expandafter\ifx\csname urlprefix\endcsname\relax\def\urlprefix{URL }\fi
\providecommand{\bibinfo}[2]{#2}
\providecommand{\eprint}[2][]{\url{#2}}

\bibitem[{\citenamefont{Chabanat et~al.}(1997)\citenamefont{Chabanat, Bonche,
  Haensel, Meyer, and Schaeffer}}]{Chabanat97}
\bibinfo{author}{\bibfnamefont{E.}~\bibnamefont{Chabanat}},
  \bibinfo{author}{\bibfnamefont{P.}~\bibnamefont{Bonche}},
  \bibinfo{author}{\bibfnamefont{P.}~\bibnamefont{Haensel}},
  \bibinfo{author}{\bibfnamefont{J.}~\bibnamefont{Meyer}}, \bibnamefont{and}
  \bibinfo{author}{\bibfnamefont{R.}~\bibnamefont{Schaeffer}},
  \bibinfo{journal}{Nucl. Phys.} \textbf{\bibinfo{volume}{A627}},
  \bibinfo{pages}{710} (\bibinfo{year}{1997}).

\bibitem[{\citenamefont{Stone et~al.}(2003)\citenamefont{Stone, Miller,
  Koncewicz, Stevenson, and Strayer}}]{Stone03}
\bibinfo{author}{\bibfnamefont{J.~R.} \bibnamefont{Stone}},
  \bibinfo{author}{\bibfnamefont{J.~C.} \bibnamefont{Miller}},
  \bibinfo{author}{\bibfnamefont{R.}~\bibnamefont{Koncewicz}},
  \bibinfo{author}{\bibfnamefont{P.~D.} \bibnamefont{Stevenson}},
  \bibnamefont{and} \bibinfo{author}{\bibfnamefont{M.~R.}
  \bibnamefont{Strayer}}, \bibinfo{journal}{Phys. Rev. C}
  \textbf{\bibinfo{volume}{68}}, \bibinfo{pages}{034324}
  (\bibinfo{year}{2003}).

\bibitem[{\citenamefont{Agrawal et~al.}(2006)\citenamefont{Agrawal, Dhiman, and
  Kumar}}]{Agrawal06}
\bibinfo{author}{\bibfnamefont{B.~K.} \bibnamefont{Agrawal}},
  \bibinfo{author}{\bibfnamefont{S.~K.} \bibnamefont{Dhiman}},
  \bibnamefont{and} \bibinfo{author}{\bibfnamefont{R.}~\bibnamefont{Kumar}},
  \bibinfo{journal}{Phys. Rev. C} \textbf{\bibinfo{volume}{73}},
  \bibinfo{pages}{034319} (\bibinfo{year}{2006}).

\bibitem[{\citenamefont{Steiner et~al.}(2005)\citenamefont{Steiner, Prakash,
  Lattimer, and Ellis}}]{Steiner05}
\bibinfo{author}{\bibfnamefont{A.~W.} \bibnamefont{Steiner}},
  \bibinfo{author}{\bibfnamefont{M.}~\bibnamefont{Prakash}},
  \bibinfo{author}{\bibfnamefont{J.~M.} \bibnamefont{Lattimer}},
  \bibnamefont{and} \bibinfo{author}{\bibfnamefont{P.}~\bibnamefont{Ellis}},
  \bibinfo{journal}{Phys. Rep.} \textbf{\bibinfo{volume}{411}},
  \bibinfo{pages}{325} (\bibinfo{year}{2005}).

\bibitem[{\citenamefont{Walecka}(1974)}]{Walecka74}
\bibinfo{author}{\bibfnamefont{J.~D.} \bibnamefont{Walecka}},
  \bibinfo{journal}{Ann. Phys. (N.Y.)} \textbf{\bibinfo{volume}{83}},
  \bibinfo{pages}{491} (\bibinfo{year}{1974}).

\bibitem[{\citenamefont{Ring}(1996)}]{Ring96}
\bibinfo{author}{\bibfnamefont{P.}~\bibnamefont{Ring}}, \bibinfo{journal}{Prog.
  Part. Nucl. Phys.} \textbf{\bibinfo{volume}{37}}, \bibinfo{pages}{193}
  (\bibinfo{year}{1996}).

\bibitem[{\citenamefont{Meng and Ring}(1996)}]{Meng96}
\bibinfo{author}{\bibfnamefont{J.}~\bibnamefont{Meng}} \bibnamefont{and}
  \bibinfo{author}{\bibfnamefont{P.}~\bibnamefont{Ring}},
  \bibinfo{journal}{Phys.Rev. Lett.} \textbf{\bibinfo{volume}{77}},
  \bibinfo{pages}{3963} (\bibinfo{year}{1996}).

\bibitem[{\citenamefont{Chen et~al.}(2005)\citenamefont{Chen, Ko, and
  Li}}]{Li2005}
\bibinfo{author}{\bibfnamefont{L.~W.} \bibnamefont{Chen}},
  \bibinfo{author}{\bibfnamefont{C.~M.} \bibnamefont{Ko}}, \bibnamefont{and}
  \bibinfo{author}{\bibfnamefont{B.~A.} \bibnamefont{Li}},
  \bibinfo{journal}{Phys. Rev.C} \textbf{\bibinfo{volume}{72}},
  \bibinfo{pages}{064309} (\bibinfo{year}{2005}).

\bibitem[{\citenamefont{Chen et~al.}(2007)\citenamefont{Chen, Ko, and
  Li}}]{Li2007}
\bibinfo{author}{\bibfnamefont{L.~W.} \bibnamefont{Chen}},
  \bibinfo{author}{\bibfnamefont{C.~M.} \bibnamefont{Ko}}, \bibnamefont{and}
  \bibinfo{author}{\bibfnamefont{B.~A.} \bibnamefont{Li}},
  \bibinfo{journal}{Phys. Rev.C} \textbf{\bibinfo{volume}{76}},
  \bibinfo{pages}{054316} (\bibinfo{year}{2007}).

\bibitem[{\citenamefont{Li et~al.}(2008)\citenamefont{Li, Chen, and Ko}}]{Li08}
\bibinfo{author}{\bibfnamefont{B.~A.} \bibnamefont{Li}},
  \bibinfo{author}{\bibfnamefont{L.~W.} \bibnamefont{Chen}}, \bibnamefont{and}
  \bibinfo{author}{\bibfnamefont{C.~M.} \bibnamefont{Ko}},
  \bibinfo{journal}{Phys. Rep.} \textbf{\bibinfo{volume}{464}},
  \bibinfo{pages}{113} (\bibinfo{year}{2008}).

\bibitem[{\citenamefont{Li et~al.}(2007)\citenamefont{Li, Garg, Liu, Marks,
  Nayak, Madhusudhana, Fujiwara, Hashimoto, Nakanishi, Okumura et~al.}}]{Li07}
\bibinfo{author}{\bibfnamefont{T.}~\bibnamefont{Li}},
  \bibinfo{author}{\bibfnamefont{U.}~\bibnamefont{Garg}},
  \bibinfo{author}{\bibfnamefont{Y.}~\bibnamefont{Liu}},
  \bibinfo{author}{\bibfnamefont{R.}~\bibnamefont{Marks}},
  \bibinfo{author}{\bibfnamefont{B.~K.} \bibnamefont{Nayak}},
  \bibinfo{author}{\bibfnamefont{P.~V.~R.} \bibnamefont{Madhusudhana}},
  \bibinfo{author}{\bibfnamefont{M.}~\bibnamefont{Fujiwara}},
  \bibinfo{author}{\bibfnamefont{H.}~\bibnamefont{Hashimoto}},
  \bibinfo{author}{\bibfnamefont{K.~K.~K.} \bibnamefont{Nakanishi}},
  \bibinfo{author}{\bibfnamefont{S.}~\bibnamefont{Okumura}},
  \bibnamefont{et~al.}, \bibinfo{journal}{Phys.Rev.Lett.}
  \textbf{\bibinfo{volume}{99}}, \bibinfo{pages}{162503}
  (\bibinfo{year}{2007}).

\bibitem[{\citenamefont{Youngblood et~al.}(1999)\citenamefont{Youngblood,
  Clark, and Lui}}]{Youngblood99}
\bibinfo{author}{\bibfnamefont{D.~H.} \bibnamefont{Youngblood}},
  \bibinfo{author}{\bibfnamefont{H.~L.} \bibnamefont{Clark}}, \bibnamefont{and}
  \bibinfo{author}{\bibfnamefont{Y.~W.} \bibnamefont{Lui}},
  \bibinfo{journal}{Phys. Rev. Lett.} \textbf{\bibinfo{volume}{82}},
  \bibinfo{pages}{691} (\bibinfo{year}{1999}).

\bibitem[{\citenamefont{Ma et~al.}(2002)\citenamefont{Ma, Wandelt, Giai, Ring,
  and Cao}}]{Ma02}
\bibinfo{author}{\bibfnamefont{Z.~Y.} \bibnamefont{Ma}},
  \bibinfo{author}{\bibfnamefont{A.}~\bibnamefont{Wandelt}},
  \bibinfo{author}{\bibfnamefont{N.~V.} \bibnamefont{Giai}},
  \bibinfo{author}{\bibfnamefont{D.~P.} \bibnamefont{Ring}}, \bibnamefont{and}
  \bibinfo{author}{\bibfnamefont{L.~G.} \bibnamefont{Cao}},
  \bibinfo{journal}{Nucl.Phys.A} \textbf{\bibinfo{volume}{703}},
  \bibinfo{pages}{222} (\bibinfo{year}{2002}).

\bibitem[{\citenamefont{Shetty et~al.}(2007)\citenamefont{Shetty, Yennello, and
  Souliotis}}]{Shetty07}
\bibinfo{author}{\bibfnamefont{D.~V.} \bibnamefont{Shetty}},
  \bibinfo{author}{\bibfnamefont{S.~J.} \bibnamefont{Yennello}},
  \bibnamefont{and} \bibinfo{author}{\bibfnamefont{G.~A.}
  \bibnamefont{Souliotis}}, \bibinfo{journal}{Phys. Rev. C}
  \textbf{\bibinfo{volume}{75}}, \bibinfo{pages}{034602}
  (\bibinfo{year}{2007}).

\bibitem[{\citenamefont{Natowitz et~al.}(2010)\citenamefont{Natowitz, Ropke,
  Typel, Blaschke, Bonasera, Hagel, Klahn, Kowalski, Qin, Shlomo
  et~al.}}]{Natowitz10}
\bibinfo{author}{\bibfnamefont{J.~B.} \bibnamefont{Natowitz}},
  \bibinfo{author}{\bibfnamefont{G.}~\bibnamefont{Ropke}},
  \bibinfo{author}{\bibfnamefont{S.}~\bibnamefont{Typel}},
  \bibinfo{author}{\bibfnamefont{D.}~\bibnamefont{Blaschke}},
  \bibinfo{author}{\bibfnamefont{A.}~\bibnamefont{Bonasera}},
  \bibinfo{author}{\bibfnamefont{K.}~\bibnamefont{Hagel}},
  \bibinfo{author}{\bibfnamefont{T.}~\bibnamefont{Klahn}},
  \bibinfo{author}{\bibfnamefont{S.}~\bibnamefont{Kowalski}},
  \bibinfo{author}{\bibfnamefont{L.}~\bibnamefont{Qin}},
  \bibinfo{author}{\bibfnamefont{S.}~\bibnamefont{Shlomo}},
  \bibnamefont{et~al.}, \bibinfo{journal}{Phys. Rev. Lett.}
  \textbf{\bibinfo{volume}{104}}, \bibinfo{pages}{202501}
  (\bibinfo{year}{2010}).

\bibitem[{\citenamefont{Tsang et~al.}(2009)\citenamefont{Tsang, Zhang,
  Danielewicz, Famiano, Li, Lynch, and Steiner}}]{Tsang09}
\bibinfo{author}{\bibfnamefont{M.~B.} \bibnamefont{Tsang}},
  \bibinfo{author}{\bibfnamefont{Y.}~\bibnamefont{Zhang}},
  \bibinfo{author}{\bibfnamefont{P.}~\bibnamefont{Danielewicz}},
  \bibinfo{author}{\bibfnamefont{M.}~\bibnamefont{Famiano}},
  \bibinfo{author}{\bibfnamefont{Z.}~\bibnamefont{Li}},
  \bibinfo{author}{\bibfnamefont{W.~G.} \bibnamefont{Lynch}}, \bibnamefont{and}
  \bibinfo{author}{\bibfnamefont{A.~W.} \bibnamefont{Steiner}},
  \bibinfo{journal}{Phys.Rev.Lett.} \textbf{\bibinfo{volume}{102}},
  \bibinfo{pages}{122701} (\bibinfo{year}{2009}).

\bibitem[{\citenamefont{Yoshida et~al.}(1998)\citenamefont{Yoshida, Sagawa, and
  Takigawa}}]{Yoshida98}
\bibinfo{author}{\bibfnamefont{S.}~\bibnamefont{Yoshida}},
  \bibinfo{author}{\bibfnamefont{H.}~\bibnamefont{Sagawa}}, \bibnamefont{and}
  \bibinfo{author}{\bibfnamefont{N.}~\bibnamefont{Takigawa}},
  \bibinfo{journal}{Phys. Rev.C} \textbf{\bibinfo{volume}{58}},
  \bibinfo{pages}{2796} (\bibinfo{year}{1998}).

\bibitem[{\citenamefont{Li et~al.}(2010)\citenamefont{Li, , Garg, Liu, Marks,
  Nayak, MadhusudhanaRao, Fujiwara, Hashimoto, Nakanishi et~al.}}]{Li10}
\bibinfo{author}{\bibfnamefont{T.}~\bibnamefont{Li}}, ,
  \bibinfo{author}{\bibfnamefont{U.}~\bibnamefont{Garg}},
  \bibinfo{author}{\bibfnamefont{Y.}~\bibnamefont{Liu}},
  \bibinfo{author}{\bibfnamefont{R.}~\bibnamefont{Marks}},
  \bibinfo{author}{\bibfnamefont{B.~K.} \bibnamefont{Nayak}},
  \bibinfo{author}{\bibfnamefont{P.~V.} \bibnamefont{MadhusudhanaRao}},
  \bibinfo{author}{\bibfnamefont{M.}~\bibnamefont{Fujiwara}},
  \bibinfo{author}{\bibfnamefont{H.}~\bibnamefont{Hashimoto}},
  \bibinfo{author}{\bibfnamefont{K.}~\bibnamefont{Nakanishi}},
  \bibnamefont{et~al.}, \bibinfo{journal}{Phys. Rev. C}
  \textbf{\bibinfo{volume}{81}}, \bibinfo{pages}{034309}
  (\bibinfo{year}{2010}).

\bibitem[{\citenamefont{Shen}(2002)}]{Shen02}
\bibinfo{author}{\bibfnamefont{H.}~\bibnamefont{Shen}}, \bibinfo{journal}{Phys.
  Rev. C} \textbf{\bibinfo{volume}{65}}, \bibinfo{pages}{035802}
  (\bibinfo{year}{2002}).

\bibitem[{\citenamefont{Schulze et~al.}(2006)\citenamefont{Schulze, Polls,
  Ramos, and Vidana}}]{Schulze06}
\bibinfo{author}{\bibfnamefont{H.~J.} \bibnamefont{Schulze}},
  \bibinfo{author}{\bibfnamefont{A.}~\bibnamefont{Polls}},
  \bibinfo{author}{\bibfnamefont{A.}~\bibnamefont{Ramos}}, \bibnamefont{and}
  \bibinfo{author}{\bibfnamefont{I.}~\bibnamefont{Vidana}},
  \bibinfo{journal}{Phys. Rev. C} \textbf{\bibinfo{volume}{73}},
  \bibinfo{pages}{058801} (\bibinfo{year}{2006}).

\bibitem[{\citenamefont{Shapiro and Teukolsky}(1983)}]{Shapiro83}
\bibinfo{author}{\bibfnamefont{S.~L.} \bibnamefont{Shapiro}} \bibnamefont{and}
  \bibinfo{author}{\bibfnamefont{S.~A.} \bibnamefont{Teukolsky}},
  \emph{\bibinfo{title}{Black Holes, White Dwarfs and Neutron Stars}}
  (\bibinfo{publisher}{Wiley, New York}, \bibinfo{year}{1983}).

\bibitem[{\citenamefont{Glendenning}(2000)}]{Glendenning00}
\bibinfo{author}{\bibfnamefont{N.~K.} \bibnamefont{Glendenning}},
  \emph{\bibinfo{title}{Compect Stars: Nuclear Physics, Particle Physics, and
  General Relativity}} (\bibinfo{publisher}{Springer-Verlag, New York},
  \bibinfo{year}{2000}).

\bibitem[{\citenamefont{Balberg et~al.}(1999)\citenamefont{Balberg,
  Lichtenstadt, and B.Cook}}]{Balberg99}
\bibinfo{author}{\bibfnamefont{S.}~\bibnamefont{Balberg}},
  \bibinfo{author}{\bibfnamefont{I.}~\bibnamefont{Lichtenstadt}},
  \bibnamefont{and} \bibinfo{author}{\bibfnamefont{G.}~\bibnamefont{B.Cook}},
  \bibinfo{journal}{Astrophys. J. Supplement} \textbf{\bibinfo{volume}{121}},
  \bibinfo{pages}{515} (\bibinfo{year}{1999}).

\bibitem[{\citenamefont{Dhiman et~al.}(2007)\citenamefont{Dhiman, Kumar, and
  Agrawal}}]{Dhiman07}
\bibinfo{author}{\bibfnamefont{S.~K.} \bibnamefont{Dhiman}},
  \bibinfo{author}{\bibfnamefont{R.}~\bibnamefont{Kumar}}, \bibnamefont{and}
  \bibinfo{author}{\bibfnamefont{B.~K.} \bibnamefont{Agrawal}},
  \bibinfo{journal}{Phys. Rev. C} \textbf{\bibinfo{volume}{76}},
  \bibinfo{pages}{045801} (\bibinfo{year}{2007}).

\bibitem[{\citenamefont{Serot and Walecka}(1997)}]{Serot97}
\bibinfo{author}{\bibfnamefont{B.~D.} \bibnamefont{Serot}} \bibnamefont{and}
  \bibinfo{author}{\bibfnamefont{J.~D.} \bibnamefont{Walecka}},
  \bibinfo{journal}{Int. J. Mod. Phys. E} \textbf{\bibinfo{volume}{6}},
  \bibinfo{pages}{515} (\bibinfo{year}{1997}).

\bibitem[{\citenamefont{Agrawal}(2010)}]{Agrawal10}
\bibinfo{author}{\bibfnamefont{B.~K.} \bibnamefont{Agrawal}},
  \bibinfo{journal}{Phys. Rev. C} \textbf{\bibinfo{volume}{81}},
  \bibinfo{pages}{034323} (\bibinfo{year}{2010}).

\bibitem[{\citenamefont{Lalazissis et~al.}(1997)\citenamefont{Lalazissis,
  Konig, and Ring}}]{Lalazissis97}
\bibinfo{author}{\bibfnamefont{G.~A.} \bibnamefont{Lalazissis}},
  \bibinfo{author}{\bibfnamefont{J.}~\bibnamefont{Konig}}, \bibnamefont{and}
  \bibinfo{author}{\bibfnamefont{P.}~\bibnamefont{Ring}},
  \bibinfo{journal}{Phys. Rev. C} \textbf{\bibinfo{volume}{55}},
  \bibinfo{pages}{540} (\bibinfo{year}{1997}).

\bibitem[{\citenamefont{Y.Sugahara and H.Toki}(1994)}]{Sugahara94}
\bibinfo{author}{\bibnamefont{Y.Sugahara}} \bibnamefont{and}
  \bibinfo{author}{\bibnamefont{H.Toki}}, \bibinfo{journal}{Nucl. Phys.}
  \textbf{\bibinfo{volume}{A579}}, \bibinfo{pages}{557} (\bibinfo{year}{1994}).

\bibitem[{\citenamefont{Baym et~al.}(1971)\citenamefont{Baym, Pethick, and
  Sutherland}}]{Baym71}
\bibinfo{author}{\bibfnamefont{G.}~\bibnamefont{Baym}},
  \bibinfo{author}{\bibfnamefont{C.}~\bibnamefont{Pethick}}, \bibnamefont{and}
  \bibinfo{author}{\bibfnamefont{P.}~\bibnamefont{Sutherland}},
  \bibinfo{journal}{Astrophys. J.} \textbf{\bibinfo{volume}{170}},
  \bibinfo{pages}{299} (\bibinfo{year}{1971}).

\bibitem[{\citenamefont{Weinberg}(1972)}]{Weinberg72}
\bibinfo{author}{\bibfnamefont{S.}~\bibnamefont{Weinberg}},
  \emph{\bibinfo{title}{Gravitation and Cosmology}} (\bibinfo{publisher}{Wiley,
  New York}, \bibinfo{year}{1972}).

\bibitem[{\citenamefont{Demorest et~al.}(2010)\citenamefont{Demorest, Pennucci,
  Ransom, Roberts, and Hessels}}]{Demorest10}
\bibinfo{author}{\bibfnamefont{P.~B.} \bibnamefont{Demorest}},
  \bibinfo{author}{\bibfnamefont{T.}~\bibnamefont{Pennucci}},
  \bibinfo{author}{\bibfnamefont{S.~M.} \bibnamefont{Ransom}},
  \bibinfo{author}{\bibfnamefont{M.~S.~E.} \bibnamefont{Roberts}},
  \bibnamefont{and} \bibinfo{author}{\bibfnamefont{J.~W.~T.}
  \bibnamefont{Hessels}}, \bibinfo{journal}{Nature}
  \textbf{\bibinfo{volume}{467}}, \bibinfo{pages}{1081} (\bibinfo{year}{2010}).

\bibitem[{\citenamefont{Dhiman et~al.}(2010)\citenamefont{Dhiman, Mahajan, and
  Agrawal}}]{Dhiman10}
\bibinfo{author}{\bibfnamefont{S.~K.} \bibnamefont{Dhiman}},
  \bibinfo{author}{\bibfnamefont{G.}~\bibnamefont{Mahajan}}, \bibnamefont{and}
  \bibinfo{author}{\bibfnamefont{B.~K.} \bibnamefont{Agrawal}},
  \bibinfo{journal}{Nucl. Phys. A} \textbf{\bibinfo{volume}{836}},
  \bibinfo{pages}{183} (\bibinfo{year}{2010}).

\bibitem[{\citenamefont{Stergioulas}(2003)}]{Stergioulas03}
\bibinfo{author}{\bibfnamefont{N.}~\bibnamefont{Stergioulas}},
  \bibinfo{journal}{Living Rev. Rel.} \textbf{\bibinfo{volume}{6}},
  \bibinfo{pages}{3} (\bibinfo{year}{2003}).

\bibitem[{\citenamefont{Stergioulas and Friedman}(1995)}]{Stergioulas95}
\bibinfo{author}{\bibfnamefont{N.}~\bibnamefont{Stergioulas}} \bibnamefont{and}
  \bibinfo{author}{\bibfnamefont{J.~L.} \bibnamefont{Friedman}},
  \bibinfo{journal}{Astrophys. J.} \textbf{\bibinfo{volume}{444}},
  \bibinfo{pages}{306} (\bibinfo{year}{1995}).

\bibitem[{\citenamefont{Goussard et~al.}(1997)\citenamefont{Goussard, Haensel,
  and Zdunik}}]{Goussard97}
\bibinfo{author}{\bibfnamefont{J.~O.} \bibnamefont{Goussard}},
  \bibinfo{author}{\bibfnamefont{P.}~\bibnamefont{Haensel}}, \bibnamefont{and}
  \bibinfo{author}{\bibfnamefont{J.~L.} \bibnamefont{Zdunik}},
  \bibinfo{journal}{Astron.Astrophys.} \textbf{\bibinfo{volume}{321}},
  \bibinfo{pages}{822} (\bibinfo{year}{1997}).

\end{thebibliography}

\newpage
\begin{table}
\caption{\label{tab1} The bulk properties of the nuclear matter at
the saturation density ($\rho_0$) for the different temperatures,
saturation density ($\rho_0$), energy per nucleon (E/A),
incompressibility coefficient for symmetric nuclear matter (K),
symmetry energy ($E_{sym}(\rho_0)$), linear density dependence of
symmetry energy (L) and effective nucleon mass/nucleon mass
($M^*_N/M_N$)}
\begin{tabular}{cccccccccc}
\hline
Force&$\zeta$&$\Delta$r &T &$\rho_{0}$&E/A&K&$E_{sym}(\rho_0)$&L&$M^*_N/M_N$\\
&&(fm)&MeV&($fm^{-3}$)&(MeV)&(MeV)&(MeV)&(MeV)&\\
\hline
BSR1&0.00&0.16&0&0.1481&-16.0192&240.0477&30.9841&59.6144&0.6052\\
&&&10&0.1481&-16.0192&240.0477&30.9841&59.6144&0.6052\\
&&&20&0.1481&-16.0194&239.7282&30.9838&59.6240&0.6052\\
&&&30&0.1464&-16.0483&211.0066&30.7468&59.3789&0.6088\\ \cline{1-10}
BSR7&0.00&0.28&0&0.1493&-16.1753&231.8574&36.9894&98.7838&0.6014\\
&&&10&0.1493&-16.1753&231.8574&36.9894&98.7838&0.6014\\
&&&20&0.1493&-16.1755&231.5692&36.9890&98.7972&0.6014\\
&&&30&0.1478&-16.2010&204.8214&36.6289&98.4333&0.6048\\
\hline
BSR8&0.03&0.16&0&0.1469&-16.0351&230.8656&31.0094&60.3747&0.6059\\
&&&10&0.1469&-16.0351&230.8656&31.0094&60.3747&0.6059\\
&&&20&0.1469&-16.0353&230.5972&31.0092&60.3831&0.6059\\
&&&30&0.1455&-16.0643&207.6021&30.8013&60.2005&0.6087\\ \cline{1-10}
BSR14&0.03&0.28&0&0.1474&-16.1838&235.4955&36.0527&93.4748&0.6078\\
&&&10&0.1474&-16.1838&235.4955&36.0527&93.4748&0.6078\\
&&&20& 0.1474&-16.1840&235.2218&36.0523&93.4878&0.6078\\
&&&30&0.1459& -16.2150&212.2077&35.7080&93.2011&0.6106\\
\hline
BSR15&0.06&0.16&0&0.1456&-16.0320&226.9275&30.9177&61.8943&0.6075\\
&&&10&0.1456&-16.0320&226.9275&30.9177&61.8943&0.6075\\
&&&20&0.1455&-16.0322&226.4806&30.9002&61.8556&0.6077\\
&&&30&0.1442&-16.0621&209.2170&30.7094&61.7048&0.6098\\ \cline{1-10}
BSR21&0.06&0.28&0&0.1452&-16.1235&220.4414&35.7123&92.5457&0.6017\\
&&&10&0.1452&-16.1235&220.4414&35.7123&92.5457&0.6017\\
&&&20&0.1451&-16.1236&220.0422&35.6861&92.4886&0.6019\\
&&&30&0.1441&-16.1486&205.8949&35.4542&92.3135&0.6036\\
\hline
\end{tabular}
\end{table}
\begin{table}
\caption{\label{tab2}
The values of maximum gravitational
mass $M_{max}$, radius
$R_{max}$, radius $R_{1.4}$ corresponding to canonical mass
 $1.4M_\odot$, and the gravitational redshift of the photon emitted
from the surface of the compact star at maximum mass $Z_{max}$ and
at canonical mass $Z_{1.4}$ for different values of the neutron
skin thickness $\Delta$r at different temperatures for the
$\omega$-meson self-coupling $\zeta$ = 0.0 with and without
hyperons. The hyperon meson coupling is $X_{\omega y}$ = 0.50.}
\begin{tabular}{ccccccccccccc}
\hline
&&& \multicolumn{5}{c}{without hyperons}&
\multicolumn{5}{c}{with hyperons}\\
\hline
Force&$\Delta$r &T &$M_{max}$ & $R_{max}$ & $R_{1.4}$ &$Z_{max}$ &$Z_{1.4}$&$M_{max}$ & $R_{max}$ & $R_{1.4}$ &$Z_{max}$ &$Z_{1.4}$\\
&(fm)&(MeV)&($M_\odot$)&(km)&(km)&&&($M_\odot$)&(km)&(km)&&\\
\hline
BSR1&0.16&0&2.43 &  11.74&   12.37 & 0.61 &   0.23&1.81&11.87&13.64&0.35&0.20\\
&&3&    2.65 & 13.11 &  14.12&  0.58  &  0.19&2.14 & 12.53 & 14.49 &0.42&  0.18\\
&&5&    2.66 &  13.21  & 14.49 &  0.57 &   0.18 &2.16  &12.82& 14.89  & 0.42 &  0.18\\
&&10&    2.78&   14.61 &  18.82  &  0.51&    0.13 &2.28 & 14.81 & 19.30 & 0.35&   0.13\\
\hline
BSR3&0.20&0&2.33 &  11.79  & 13.48&    0.55  &  0.20& 1.73&   11.65&   13.60 &   0.33&   0.20\\
&&3&    2.59 &  13.43  & 15.62    &    0.52  &  0.17 &2.13&   13.16&   16.03&     0.38 &   0.16\\
&&5& 2.59&   13.46  & 15.72  &    0.52  &  0.16 &2.14&   13.25 & 16.16 &  0.38 &   0.16\\
&&10&    2.65 &  14.11  & 17.06 &      0.50&    0.15&2.23 &  14.30&   17.56&    0.36&    0.14\\
\hline
BSR5&0.24&0&2.45 &  12.11&   13.75&  0.58   & 0.20 & 1.82 &  11.80&   13.88&   0.35&    0.19\\
&&3&    2.73 &  13.83 &  16.00 &    0.55  &  0.16&2.15&  13.26&  16.19&  0.39&   0.16\\
&&5&    2.73  & 13.84&   16.08 &   0.55 &   0.16 & 2.16&   13.33&   16.29& 0.38
 &   0.16\\
&&10&    2.78  & 14.30&   17.22   &  0.53&    0.15& 2.23& 14.30&   17.56&  0.36&    0.14\\
\hline
BSR7&0.28&0&2.47  & 12.23&   14.00&    0.58   & 0.19 &1.81 &  11.91 &  14.13 & 0.35   & 0.19\\
&&3&    2.80 &  14.29 &  16.57 &   0.54  &  0.15&2.19 &  13.71 &  16.69  &  0.38  &  0.15\\
&&5&    2.80&   14.30  & 16.62  &  0.54 &   0.15 &2.19 &  13.78 &  16.77  &   0.37  &  0.15\\
&&10&    2.83 &  14.60  & 17.47  & 0.53&    0.14 &2.26 & 14.59&   17.77&   0.37&   0.14\\
\hline
\end{tabular}
\end{table}
\newpage
\begin{table}
\caption{\label{tab3}
Same as Table \ref{tab2} but with $\omega$-meson self-coupling $\zeta$ = 0.03.}
\begin{tabular}{ccccccccccccc}
\hline
&&& \multicolumn{5}{c}{without hyperons}&
\multicolumn{5}{c}{with hyperons}\\
\hline
Force&$\Delta$r &T &$M_{max}$ & $R_{max}$ & $R_{1.4}$ &$Z_{max}$ &$Z_{1.4}$&$M_{max}$ & $R_{max}$ & $R_{1.4}$ &$Z_{max}$ &$Z_{1.4}$\\
&(fm)&(MeV)&($M_\odot$)&(km)&(km)&&&($M_\odot$)&(km)&(km)&&\\
\hline
BSR8&0.16&0 & 1.94 &  11.43 &  13.08 &   0.42  &  0.21&1.54&11.82&13.14&0.28&0.21 \\
&&3     & 2.18 &  12.99 &  15.07 &   0.41  &  0.17&1.84&12.92&15.50&0.31&0.17\\
&&5     & 2.19 &  13.05 &  15.21 &   0.41  &  0.17& 1.85&13.07&15.68&0.31&0.17\\
&&10    & 2.28 &  13.96 &  16.78 &   0.39  &  0.15&1.96&14.39&17.24& 0.30& 0.15\\
\hline
BSR10&0.20&0 & 1.94 &  11.46 &  13.19 &   0.41  &  0.21 & 1.54&11.80&13.24&0.28&0.21\\
&&3     & 2.21 &  13.20 &  15.43 &   0.41  &  0.17 &1.85&13.04&15.72&0.31&0.17\\
&&5     & 2.21 &  13.24 &  15.53 &   0.41  &  0.17 &1.86&13.16&15.86&0.31&0.16\\
&&10    & 2.27 &  13.91 &  16.86 &   0.39  &  0.15&1.96&14.31&17.32&0.30&0.15\\
\hline
BSR12&0.24&0 & 1.95 &  11.49 &  13.29 &   0.42  &  0.21 &1.54& 11.84&13.33&0.27&0.20\\
&&3     & 2.25 &  13.43 &  15.73 &   0.41  &  0.16&1.87&13.23&15.92&0.31& 0.16\\
&&5     & 2.25 &  13.45 &  15.80 &   0.41  &  0.16&1.88&13.31&16.03&0.31& 0.16\\
&&10    & 2.29 &  14.00 &  16.94 &   0.39  &  0.15&1.95&14.30&17.35&0.29& 0.15\\
\hline
BSR14&0.28&0 & 1.94 &  11.54 &  13.51 &   0.41  &  0.20&1.54&11.89&13.53&0.27& 0.20\\
&&3   & 2.30 &  13.82 &  16.20 &   0.40  &  0.16&1.89&13.45&16.21&0.31&0.16\\
&&5   & 2.30 &  13.84 &  16.25 &   0.40  &  0.16&1.89&13.48&16.29&0.31&0.16\\
&&10   & 2.32 &  14.25 &  17.15 &   0.39  &  0.15&1.95&14.36&17.44&0.29&0.15\\
\hline
\end{tabular}
\end{table}
\newpage
\begin{table}
\caption{\label{tab4}
Same as Table \ref{tab2} but with $\omega$-meson self-coupling $\zeta$ = 0.06.}
\begin{tabular}{ccccccccccccc}
\hline
&&& \multicolumn{5}{c}{without hyperons}&
\multicolumn{5}{c}{with hyperons}\\
\hline
Force&$\Delta$r &T &$M_{max}$ & $R_{max}$ & $R_{1.4}$ &$Z_{max}$ &$Z_{1.4}$&$M_{max}$ & $R_{max}$ & $R_{1.4}$ &$Z_{max}$ &$Z_{1.4}$\\
&(fm)&(MeV)&($M_\odot$)&(km)&(km)&&&($M_\odot$)&(km)&(km)&&\\
\hline
BSR15&0.16&0 & 1.73 &  10.92 &  12.62 &   0.37  &  0.22&1.41&11.52&12.15&0.25& 0.23\\
&&3   & 1.97 &  12.53 &  14.73 &   0.37  &  0.18 & 1.67&12.14&14.64&0.30&0.18\\
&&5   & 1.97 &  12.60 &  14.86 &   0.36  &  0.18 &1.68&12.31&14.85&0.30&0.18\\
&&10   & 2.03 &  13.41 &  16.36 &   0.35  &  0.16&1.76&13.54&16.64&0.27& 0.15\\
\hline
BSR17&0.20&0   & 1.73 &  10.93 &  12.66 &   0.37  &  0.22&1.41&11.49&12.09&0.25& 0.23\\
&&3    & 1.99 &  12.66 &  14.95 &   0.37  &  0.18 &1.69&12.33&14.88&0.30& 0.18\\
&&5    & 1.99 &  12.70 &  15.04 &   0.36  &  0.17 &1.69&12.44&15.03& 0.29& 0.17\\
&&10   & 2.04 &  13.38 &  16.40 &   0.35  &  0.16&1.76&13.56&16.71&0.27&0.15\\
\hline
BSR19&0.24&0   & 1.73 &  11.01 &  12.83 &   0.37  &  0.22&1.41&11.56&12.24&0.25& 0.23\\
&&3   & 2.03 &  12.99 &  15.41 &   0.36  &  0.17 &1.71&12.61&15.26&0.29&0.17\\
&&5   & 2.03 &  13.04 &  15.48 &   0.36  &  0.17 &1.71&12.69&15.38&0.29&0.17\\
&&10   & 2.06 &  13.56 &  16.62 &   0.35  &  0.15&1.77&13.73&16.90&0.27& 0.15\\
\hline
BSR21&0.28&0   & 1.75 &  11.17 &  13.13 &   0.36  &  0.21&1.43&11.69&12.68&0.25& 0.22\\
&&3   & 2.09 &  13.49 &  15.97 &   0.36  &  0.16 &1.75&12.96&15.77&0.29&0.16\\
&&5    & 2.09 &  13.51 &  16.02 &   0.36  &  0.16 &1.75&13.02&15.84&0.29&0.16\\
&&10    & 2.12 &  13.91 &  16.92 &   0.35  &  0.15&1.80&13.88&17.11&0.27&0.15\\
\hline
\end{tabular}
\end{table}
\newpage



\end{document}